\documentclass[final,3p,times,twocolumn,authoryear]{elsarticle}
\usepackage{amsmath, amsfonts}
\usepackage{graphicx}
\usepackage{lineno,hyperref}
\usepackage[acronym]{glossaries}
\usepackage{siunitx}
\usepackage{subcaption}
\usepackage{bm}
\usepackage{textcomp}
\usepackage{threeparttable}
\usepackage{multirow}
\usepackage{makecell}

\usepackage{url}

\makeatletter
\newcommand\mautoref[1]{\@first@ref#1,@}
\def\@throw@dot#1.#2@{#1}% discard everything after the dot
\def\@set@refname#1{%    % set \@refname to autoefname+s using \getrefbykeydefault
    \edef\@tmp{\getrefbykeydefault{#1}{anchor}{}}%
    \xdef\@tmp{\expandafter\@throw@dot\@tmp.@}%
    \ltx@IfUndefined{\@tmp autorefnameplural}%
         {\def\@refname{\@nameuse{\@tmp autorefname}s}}%
         {\def\@refname{\@nameuse{\@tmp autorefnameplural}}}%
}
\def\@first@ref#1,#2{%
  \ifx#2@\autoref{#1}\let\@nextref\@gobble% only one ref, revert to normal \autoref
  \else%
    \@set@refname{#1}%  set \@refname to autoref name
    \@refname~\ref{#1}% add autoefname and first reference
    \let\@nextref\@next@ref% push processing to \@next@ref
  \fi%
  \@nextref#2%
}
\def\@next@ref#1,#2{%
   \ifx#2@ and~\ref{#1}\let\@nextref\@gobble% at end: print and+\ref and stop
   \else, \ref{#1}% print  ,+\ref and continue
   \fi%
   \@nextref#2%
}
\makeatother

\newacronym{bdc}{BDC}{Bottom Dead Center}
\newacronym{ai}{AI}{Artificial Intelligence}
\newacronym{fcnn}{FCNN}{Fully Connected Neural Network}
\newacronym{ann}{ANN}{Artificial Neural Network}
\newacronym{mlp}{MLP}{Multilayer Perceptron}
\newacronym{cnn}{CNN}{Convolutional Neural Network}
\newacronym{cvnn}{CVNN}{Complex-Valued Neural Network}
\newacronym{ae}{AE}{Autoencoder}
\newacronym{fe}{FE}{Frequency Encoding}
\newacronym{bn}{BN}{Batch Normalization}
\newacronym{me}{ME}{Mean Error}
\newacronym{mse}{MSE}{Mean Squared Error}
\newacronym{mae}{MAE}{Mean Absolute Error}
\newacronym{rmse}{RMSE}{Root Mean Squared Error}
\newacronym{re}{RE}{Relative Error}
\newacronym{rf}{RF}{Random Forest}
\newacronym{cart}{CART}{Classification And Regression Trees}
\newacronym{gbdt}{GBDT}{Gradient Boosted Decision Tree}
\newacronym{dt}{DT}{Decision Tree}
\newacronym{svm}{SVM}{Support Vector Machine}
\newacronym{knn}{k-NN}{K-Nearest-Neighbor}

\makeglossaries

\modulolinenumbers[5]

\journal{Expert Systems with Applications}

\begin{document}

\begin{frontmatter}

%% Title, authors and addresses

%% use the tnoteref command within \title for footnotes;
%% use the tnotetext command for theassociated footnote;
%% use the fnref command within \author or \affiliation for footnotes;
%% use the fntext command for theassociated footnote;
%% use the corref command within \author for corresponding author footnotes;
%% use the cortext command for theassociated footnote;
%% use the ead command for the email address,
%% and the form \ead[url] for the home page:
%% \title{Title\tnoteref{label1}}
%% \tnotetext[label1]{}
%% \author{Name\corref{cor1}\fnref{label2}}
%% \ead{email address}
%% \ead[url]{home page}
%% \fntext[label2]{}
%% \cortext[cor1]{}
%% \affiliation{organization={},
%%            addressline={}, 
%%            city={},
%%            postcode={}, 
%%            state={},
%%            country={}}
%% \fntext[label3]{}

\title{Deep Learning-Based Position Detection for Hydraulic Cylinders Using Scattering Parameters}

%% use optional labels to link authors explicitly to addresses:
%% \author[label1,label2]{}
%% \affiliation[label1]{organization={},
%%             addressline={},
%%             city={},
%%             postcode={},
%%             state={},
%%             country={}}
%%
%% \affiliation[label2]{organization={},
%%             addressline={},
%%             city={},
%%             postcode={},
%%             state={},
%%             country={}}

\author[uni,liebherr]{Chen Xin}
\ead{chen.xin@uni-tuebingen.de}

\author[liebherr]{Thomas Motz}
\ead{thomas.motz@liebherr.com}

\author[uni]{Wolfgang Fuhl}
\ead{wolfgang.fuhl@uni-tuebingen.de}

\author[liebherr]{Andreas Hartel}
\ead{andreas.hartel@liebherr.com}

\author[tum]{Enkelejda Kasneci\corref{correspondingauthor}}
\cortext[correspondingauthor]{Corresponding author}
\ead{enkelejda.kasneci@tum.de}

\affiliation[uni]{organization={Eberhard Karls University of Tübingen},%Department and Organization
            addressline={Sand 14}, 
            city={Tübingen},
            postcode={72076}, 
            country={Germany}}

\affiliation[liebherr]{organization={Liebherr-Electronics and Drives GmbH},%Department and Organization
            addressline={Peter-Dornier-Straße 11}, 
            city={Lindau (Bodensee)},
            postcode={88131}, 
            country={Germany}}

\affiliation[tum]{organization={Technical University of Munich},%Department and Organization
            addressline={Arcisstraße 21}, 
            city={München},
            postcode={80333}, 
            country={Germany}}

\newpageafter{author}

\begin{abstract}
Position detection of hydraulic cylinder pistons is crucial for numerous industrial automation applications. A typical traditional method is to excite electromagnetic waves in the cylinder structure and analytically solve the piston position based on the scattering parameters measured by a sensor. The core of this approach is a physical model that outlines the relationship between the measured scattering parameters and the targeted piston position. However, this physical model has shortcomings in accuracy and adaptability, especially in extreme conditions. To address these limitations, we propose machine learning and deep learning-based methods to learn the relationship directly in a data-driven manner. As a result, all deep learning models in this paper consistently outperform the physical one by a large margin. We further deliberate on the choice of models based on domain knowledge and provide in-depth analyses combining model performance with real-world physical characteristics. Specifically, we use \acrfull{cnn} to discover local interactions of input among adjacent frequencies, apply \acrfull{cvnn} to exploit the complex-valued nature of electromagnetic scattering parameters, and introduce a novel technique named \acrlong{fe} to add weighted frequency information to the model input.  The combination of these techniques results in our best-performing model, a complex-valued \acrshort{cnn} with \acrlong{fe}, which exhibits substantial improvement in accuracy with an error reduction of 1/12 compared to the traditional physical model.
\end{abstract}

% %%Graphical abstract
% \begin{graphicalabstract}
% \includegraphics{grabs}
% \end{graphicalabstract}

% %%Research highlights
% \begin{highlights}
% \item Research highlight 1
% \item Research highlight 2
% \end{highlights}


\end{frontmatter}

%% \linenumbers

%% main text

\section{Introduction}

The ubiquitous aim for automation in industrial processes demands accurate control of their components. To better control the  working behavior of a hydraulic cylinder, one of the most widely used components in industrial automation applications, we need to know the precise position of its piston. LiView (\autoref{fig:liview}) is a sensor that can emit microwave signals and intercept their reflections inside a hydraulic cylinder to detect the piston position. 

One can model the physical relationship between the microwave signals and the current piston position by interpreting the cylinder as a linear electrical network to compute the exact value of the current piston position. This network is coupled to ports that are excited by the incident and reflected voltage waves that reflect the interaction of the electromagnetic waves and the cylinder\citep{white2004high}. The whole network can then be formalized by a linear equation, where its coefficients, the scattering parameters, convey the excitation of the ports and detailed model assumptions of the network. After measuring these scattering parameters, we can use the mathematical expression of the network in reverse to deduce the piston position. However, due to the high demand for positional accuracy, the calculation of piston position based on the physical model, the linear electrical network based on the scattering matrix, is problematic. One of the arguments is that the model performance deteriorates severely in harsh environments and challenging use cases because extreme conditions invalidate the physical model. Additionally, the sensor should function properly and guarantee high accuracy in a large variety of temperatures, mechanical stress, moisture, and dust, which is precisely the weakness of the approach based on the physical model, given its poor consistency and robustness. On the contrary, these requirements make the learning-based methods seem more promising. 

Machine learning is the process of using adaptive models to analyze and understand patterns in data, and make predictions without explicit programming. A typical workflow is to hand-design good features from raw data, and then the machine learning models are trained in to map the generated features to the corresponding target. However, the feature extraction from scattering parameters is a nontrivial task that requires in-depth domain expertise, The quality of the features directly affects the accuracy. We confirm this through experiments, where we apply classical machine learning models directly to the raw data and are not able to achieve satisfactory results. Due to the complexity of the environment inside the cylinder, we do not force meticulous engineering on handcrafted features but pin our hopes on deep learning instead. 

Deep learning, or \acrfull{ann}, as an important branch of machine learning, provides a way to avoid the necessity for handcrafted features. It allows models composed of multiple processing layers to learn data representations with multiple levels of abstraction automatically. Two quintessential examples of \acrshort{ann}s are \acrfull{mlp} and \acrfull{cnn}, which are multilayer stacks of fully connected and convolutional layers, respectively. In this paper, we explore the feasibility of \acrshort{mlp}s and \acrshort{cnn}s for piston position detection based on scattering parameters. In all our experiments, deep learning-based models consistently outperform both the traditional physical model and classical machine learning models by a large margin. We also investigate the particular impact of general architectures and components of \acrshort{mlp}s and \acrshort{cnn}s for this specific task. We find that different activation functions and the application of convolution have significant effects on the model performance for our specific task. We offer explanations in the context of the characteristics of the model and the physical nature of hydraulic cylinders.

The aforementioned machine learning and deep learning models mainly focus on real-valued data. However, scattering parameters are complex numbers. One way to cope with complex numbers is to break a complex-valued vector into its real and imaginary parts to form a real-valued vector of two times the length of the original complex-valued vector. This simple approach effortlessly translates the problem from the complex domain to the real domain but the essential statistical information, such as the correlation between the real and imaginary parts of the numbers is completely lost. On the other hand, \acrfull{cvnn} with complex-valued weights, such as complex-valued \acrshort{mlp} \citep{hiroseComplexValuedNeuralNetworks2012,sarroffComplexNeuralNetworks} and complex-valued \acrshort{cnn} \citep{trabelsiDeepComplexNetworks2018,zhangComplexValuedConvolutionalNeural2017}, solve the problem directly in the complex domain. They are designed to have more constraints to force the model to mimic calculations in the complex domain. This way, more information about the specific application is introduced to the model, which could improve the model performance. In this paper, we show that our \acrshort{cvnn} versions of \acrshort{mlp} and \acrshort{cnn} both substantially outperform their real-valued counterparts.

In addition to internalizing the information that the model should operate in the complex domain by \acrshort{cvnn}, other information related to the input can also be introduced into the model by simply adding the information to the input itself. For example, in the original transformer model \citep{vaswaniAttentionAllYou2017}, the position information of the input words is encoded and added directly to the word embedding. Inspired by this idea, we also applied a model-agnostic technique named \acrlong{fe} to add frequency information to our model input after weighting by learnable parameters. To our best knowledge, this is the first work to use frequency information of electromagnetic waves by adding a learned weighted encoding of it directly to the model input. This simple trick significantly improves the model performance by about 36\% for both \acrshort{mlp} and \acrshort{cnn} framework.  

Machine learning \citep{roshaniDesignModelingCompact2021,akhavanhejaziNewOnlineMonitoring2011,barbosaMachineLearningApproach2019,ranaMachineLearningApproaches,baderIdentificationCommunicationCables2021,hejaziOnlineMonitoringTransformer2011} and deep learning \citep{hienMaterialThicknessClassification2021,frazierDeepLearningEstimationComplex2022,husbyEddyCurrentDuplex,guptaHandMovementClassification,travassosArtificialNeuralNetworks2020,ranaMachineLearningApproaches} have already caught the attention of previous researchers in scattering parameter related projects. Despite the notable achievements in accuracy, there are still some regretful missing aspects. We consistently find the following pattern in papers in this area: the authors take a great deal of effort to design an elaborate physical experiment but only present a standard machine learning or deep learning model without the motivation for the choice of their specific models, the comparison between different models, and their connections to the real-world physical system. In contrast, we keep the physical model unchanged and focus on designing the best deep learning models. We find that our best deep learning model designed under the guidance of physical properties can perform 4 times better than a standard one with similar model parameters. This is strong evidence of the potential of dedicated deep learning models in the area of scattering parameter-related projects.  

In this paper, we propose promising deep learning models for piston detection using scattering parameters. We elaborate on the motivations for the chosen models, compare the performance of different learning-based models, conjecture possible model improvements based on the physical system, and prove our assumptions by in-depth analyses based on the model characteristics and physical nature of the task. We hope this paper can serve as a guide for projects in the area of deep learning using scattering parameters and a reminder of the potential of dedicated deep learning models. Our major findings and contributions are as follows:

\begin{enumerate}

    \item Deep learning-based models significantly and consistently outperform both the traditional physical model and classical machine learning models. The relative error of the most accurate model, a complex-valued \acrshort{cnn} with \acrfull{fe}, is less than 1/12 of the traditional physical model.
    \item \acrlong{fe} dramatically improves the generalization performance of all \acrshort{ann} based models at a negligible cost of extra trainable parameters and without the need to change the backbone structure of the model.
    \item \acrlong{cvnn}s outperform their real-valued counterparts by a large margin, which is strong evidence that more emphasis should be placed on the complex nature of the specific task. 
    \item \acrshort{cnn}s consistently beat \acrshort{mlp}s on performance, indicating the local interactions between frequencies play an important role.
    \item The performance of models with the same architecture, but different activation functions varies greatly, implying that a function choice that more closely matches the physical characteristics of the task is preferred, especially when computational resources limit the total number of parameters of the model.
\end{enumerate}

The rest of the paper is organized as follows. In section \ref{sec:related_work}, we summarize the achievements and our inspirations from them. Then in section \ref{sec:liview}, we show the physical characteristics by formalizing the physical model. All the machine learning (\acrfull{rf} and \acrfull{gbdt} in section \ref{sec:ml}), and deep learning models (\acrshort{mlp} and \acrshort{cnn} in section \ref{sec:dl}) including \acrshort{cvnn} (section \ref{sec:cvnn}) are presented in section \ref{sec:methods}, together with the novel technique \acrlong{fe} (section \ref{sec:fe}). In section \ref{sec:experiments}, we introduce the dataset (section \ref{sec:dataset}), specify the implementation details (section \ref{sec:implementation}), evaluate the model performance in accuracy (section \ref{sec:comparison}) and lastly discuss model complexity in terms of parameter numbers and inference speed (section \ref{sec:complexity}). Finally, section \ref{sec:conclusion} provides our conclusions.

\section{Related Work}
\label{sec:related_work}

\textbf{Physical Model-Based Approaches.}
Given the characteristics of the interior of hydraulic cylinders, electromagnetic wave is a good medium for detecting the piston position. Compared to alternatives like optical systems or cable potentiometers, a sensor based on electromagnetic waves is less exposed to external perturbations since the measurement depends solely on the cylinder's interior. 
There are some varieties of attempts to measure the piston position with electromagnetic signals. The difference lies in whether coupling the signal into the cylinder on the side of the piston where the rod is present \citep{Caterpillar1994, Braun2014, Scheidt2017} or on the other side without the rod \citep{Balluff1, Balluff2}. In the first situation, the cylinder tube and piston rod form a coaxial cavity, while the latter leads to a signal propagating in a hollow waveguide of variable length. Both variants can be modeled so that the position of the cylinder can be extracted, for example, from the analysis of resonant frequencies. 
The physical model we adopt for the comparison to our deep learning models is the LiView system (section \ref{sec:liview}). It belongs to the first category. A detailed description of it can be found in section \ref{sec:liview}. In short, we model the cylinder as a three-port electrical system (\autoref{eq:matrix_three_port}) and derive equations (see \mautoref{eq:transmission,eq:t_of_l}) describing the relationship between the target piston position and our sensor measurements. The equations are solvable only if we make some reasonable assumptions and fix the values of some variables by means of calibrations in advance. However, the variable assumptions could break in harsh environments and the model assumptions are not exactly accurate in extreme conditions, which leads to noisy and inaccurate position detections. Learning-based methods overcome these restrictions and allow for a generic modeling process for various environments.

\textbf{Machine Learning-Based Approaches.}
Classical machine learning algorithms, such as \acrfull{svm}, \acrfull{knn}, \acrfull{dt} have been widely adopted in industrial applications \citep{bertoliniMachineLearningIndustrial2021,kangMachineLearningApplications2020,narcisoApplicationMachineLearning2020}. Machine Learning has also become a popular choice to find the relationship between the input scattering parameters and the desired target. For instance, a \acrshort{knn}-based method for online monitoring of transformer winding axial displacement using scattering parameters is presented in \cite{hejaziOnlineMonitoringTransformer2011}. The closeness among samples is measured in terms of Euclidean distance in the feature space consisting of the magnitude and phase of the measured scattering parameters. Researchers in the same group used the same feature space but replace \acrshort{knn} with \acrfull{cart} \citep{reason:BreFriOlsSto84a} to detect the axial position of the winding \citep{akhavanhejaziNewOnlineMonitoring2011}. Both \acrshort{knn} and \acrshort{dt} were able to detect the position of the winding (displacement) with acceptable relative errors (smaller than 0.2\%). But this performance could only be achieved after storing enough reference feature vectors in advance. In recent studies, classical machine learning approaches are still popular. \cite{baderIdentificationCommunicationCables2021} investigated a cable identification method using \acrshort{svm} based on scattering parameters. The real and imaginary parts of the complex-valued scattering parameters and their magnitudes were used as the input of a linear \acrshort{svm} which achieved an accuracy close to 100\%. Despite the promising results, the applications mentioned above are rooted in relatively simple systems (specifically two-port networks simulated in the lab). In this paper, we used two more sophisticated machine learning models, \acrshort{rf} and \acrshort{gbdt}, to handle more complicated (three-port network) and noisy real-world experiments. But without forcing meticulous feature engineering, machine learning models do not perform as well as deep learning models.

\textbf{Deep Learning-Based Approaches.} 
Deep Learning has attracted the attention of researchers working with scattering parameters due to its strong potential both in theory and in practice. \cite{husbyEddyCurrentDuplex} adopted 3-hidden-layer \acrshort{mlp} with ReLU activation to detect the thickness of duplex coatings used for corrosion protection of carbon steel substrates. They concatenated the real and imaginary components of the measured scattering parameters into a real vector with doubled dimensions as the input to the \acrshort{mlp}. They were able to reduce the relative standard deviation to 2.5\%. \cite{hienMaterialThicknessClassification2021} proposed a material thickness classification with the help of a 6-hidden-layer \acrshort{mlp} also with ReLU activation using features derived from scattering parameters. Besides real and imaginary parts of the scattering parameters as in \cite{baderIdentificationCommunicationCables2021}, they also used frequency, relative permittivity, and loss tangent directly to form the dimensions of the input feature vectors. They binned the ground truth into eight thickness classes and recorded impeccable average estimation accuracy. Similarly,  In addition to \acrshort{mlp}, \acrshort{cnn} can also be of help. \cite{guptaHandMovementClassification} described a \acrshort{cnn} based hand movement classification from scattering parameters. Likewise, they also fed their model with extra features calculated from recorded scattering parameters and finally achieved an accuracy of 98\%. 
Given the above successful demonstrations, we also evaluated \acrshort{mlp} and \acrshort{cnn} with reasonable adaptions (section \ref{sec:dl} and \ref{sec:implementation}) to fit our task. Note that all previously mentioned models in this section are real-valued models but with complex-valued input. A common way to solve this contradiction is to concatenate a complex-valued vector's real and imaginary parts to form a double-sized real-valued vector. We adopted this way to construct our real-valued models as well, but also implemented \acrshort{cvnn}s to handle complex-valued input directly. 

\textbf{\acrfull{cvnn}}
 complex numbers are often used in many real-world practical applications, such as in telecommunications, robotics, bioinformatics, image processing, sonar, radar, and speech recognition \citep{basseySurveyComplexValuedNeural2021}. Scattering parameters are also complex numbers. To deal with them directly in the complex domain, \acrshort{cvnn} is a good option. \cite{yangLandmineDetectionClassification2005} applied \acrshort{cvnn} for landmine detection based on scattering parameters. They trained \acrshort{cvnn} with a single hidden layer on a highly unbalanced small dataset. This can only result in a good average true positive rate after strict outlier removal and data balancing. In a recent study, \cite{frazierDeepLearningEstimationComplex2022} investigated \acrshort{cvnn} more thoroughly in their application for the estimation of complex reverberant wave fields using complex-valued scattering parameter as input. They tested a \acrshort{cvnn} with complex-valued weights and a hybrid \acrshort{ann} which processes real and imaginary parts of the input independently and only combines these two parts at the very end of the model to generate the output. They compared these two versions of \acrshort{cvnn} with their real-valued counterpart and found out that the \acrshort{cvnn} with complex-valued weights achieved the best performance and outperformed the real-valued model based only on the magnitude of the scattering parameters by a large margin. Besides complex-valued \acrshort{mlp} in the above-mentioned two projects, complex-valued \acrshort{cnn}s also achieved promising results in other areas, such as image recognition in \cite{trabelsiDeepComplexNetworks2018}. Encouraged by these results, we also implemented both complex-valued \acrshort{mlp} and complex-valued \acrshort{cnn} for our project and show that they outperform their real-valued counterparts by a large margin.

\textbf{Frequency Encoding.} In the aforementioned work, it's common to use not only scattering parameters but also other extra useful features either deriving from scattering parameters or related to the experimental setup, such as voltage standing wave ratio in \cite{guptaHandMovementClassification} or permittivity in \cite{hienMaterialThicknessClassification2021}. However, these extra features, although directly related to the original input, are usually concatenated to the original input, which increases the input dimension and therefore the model size significantly. Another way to introduce additional information to the model is to add extra information to the original input. A typical example is the positional encoding for the Transformer \citep{vaswaniAttentionAllYou2017}, where sinusoidal functions encode the position of each word (of the input sentence), and the code is directly added to the input embedding. Inspired by the idea of positional encoding, we designed a novel technique called \acrlong{fe}, which inserts the frequency information into the model in an additive (rather than concatenative as in previous works) way. The difference between positional encoding and our \acrlong{fe} is that we do not encode the frequency in a predefined way but with learnable parameters.

Despite the notable achievements of the papers in the area of machine learning and deep learning based on scattering parameters, there are still some regretful missing parts. Firstly, the computational resource doesn't seem to be a concern for most of the work. Considering the model deployment in an edge computing device in the future, we restrict the model parameters to 10K and provide detailed analyses of the tradeoff between model performance and model size. Secondly, previous works in this area tend to directly adopt standard deep learning models without customization based on the task at hand. The researchers focus more on the design of physical experiments but usually simply pick a standard deep learning model without the analysis of their specific models and the comparison between different models. We are often unable to find a detailed analysis of the test results, not to mention linking the model performance to practice by explaining why the model works in view of both the properties of the chosen machine learning or deep learning models and the physical characteristics of the system. In our opinion, the motivation and analyses serve as helpful guidance for similar projects in the future. Therefore, we focus on the design of the deep learning model in combination with the physical characteristics of the system. We provide in-depth analyses of the motivation for our model choices (section \ref{sec:methods}), and the evaluation of our models' performance (section \ref{sec:comparison}). But before all that, we will start our journey by introducing our physical model.

\section{Physical Model}
\label{sec:liview}

The physical model is based on the LiView system. LiView is a sensor that measures the current position of the piston in a hydraulic cylinder via the analysis of electrical fields that are excited within the cylinder \citep{Scheidt2017}. It feeds microwave signals with frequencies in a range from \SI{300}{\mega\hertz} to \SI{1.5}{\giga\hertz} into the cylinder and measures the signals reflected by the cylinder structure. The measured data provide signatures that allow a determination of the piston position after the read-out of them. 

\begin{figure}[h]
	\centering
	\begin{subfigure}[b]{0.8\columnwidth}
		\centering
		\includegraphics[width=1\linewidth]{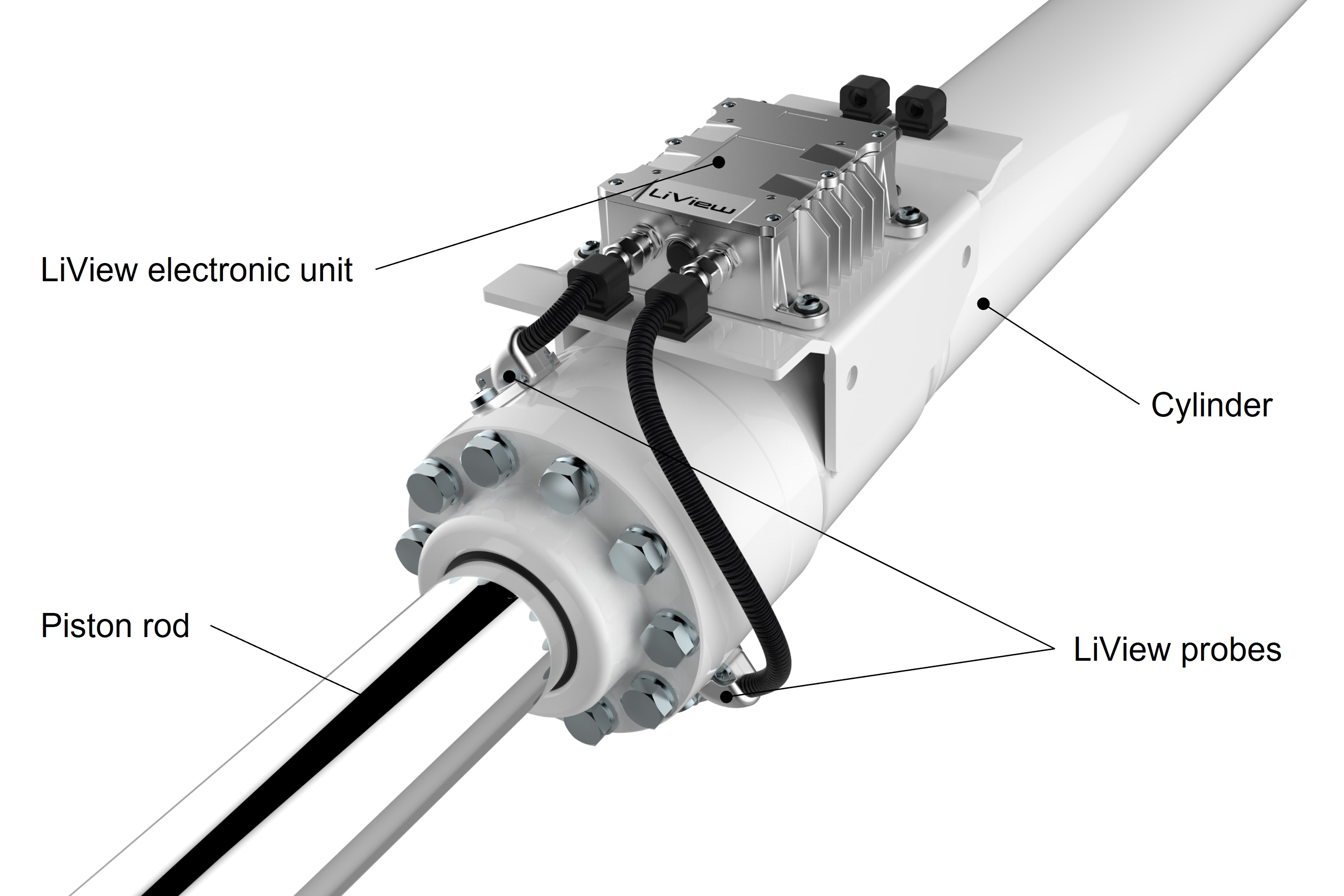}
		\caption{LiView device outside the cylinder}
		\label{fig:liview_ext}
	\end{subfigure}
	\begin{subfigure}[b]{0.9\columnwidth}
		\centering
		\includegraphics[width=1\linewidth]{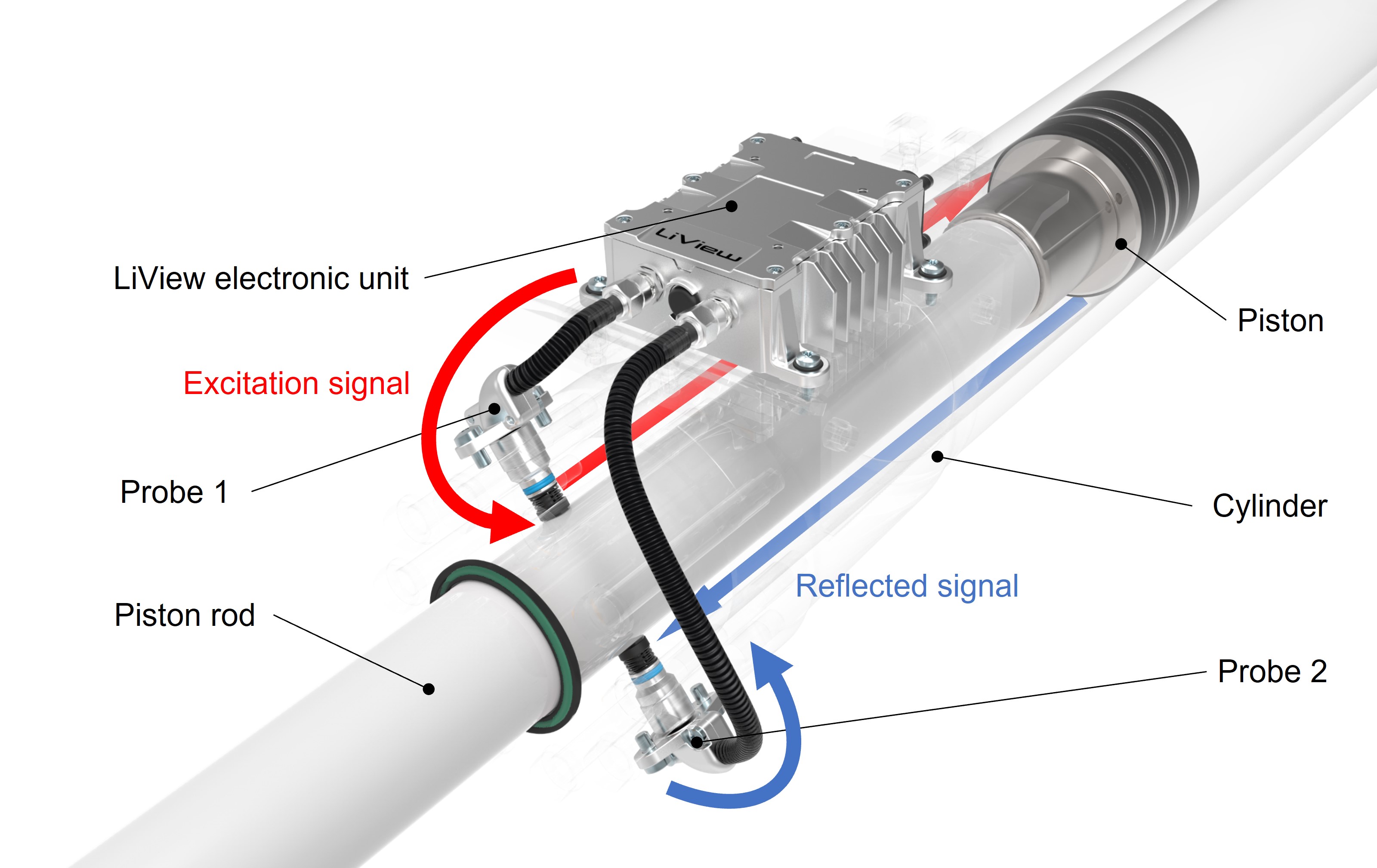}
		\caption{LiView signal inside the cylinder}
		\label{fig:liview_int}
	\end{subfigure}
	
	\caption{LiView device and signal. (a) illustrates a LiView device mounted outside a cylinder, with its probes connected to the cylinder structure. (b) shows the process of a measurement cycle of the LiView system, where the excitation and reflected signals are depicted as red and blue arrows, respectively. The LiView electronic unit first generates a microwave excitation signal, which is conducted through the black wire to probe 1 and capacitively fed into the cylinder there (the curved red arrow). The signal then propagates inside the oil between the piston rod and the cylinder until it hits the piston (the straight red arrow). Then, the signal is reflected by the piston and can be intercepted by probe 2 (the straight blue arrow). Finally, the signal is received by the LiView electronic unit (the curved blue arrow) and is processed there into the final measurement.}
	\label{fig:liview}
\end{figure}

 As shown in \autoref{fig:liview_ext}, the LiView system consists of an electronic unit positioned outside the cylinder and two probes connecting the cylinder's interior with the electronic unit via two coaxial wires. These wires conduct a microwave signal (illustrated as arrows in \autoref{fig:liview_int}) that is generated in the electronic unit and capacitively fed into the cylinder \citep{Leutenegger2017}. The electrical fields that propagate through the cylinder have frequencies in a range from \SI{300}{\mega\hertz} to \SI{1.5}{\giga\hertz} with their phases and amplitudes depending on the piston position. 

\begin{figure}
    \centering
    \includegraphics[width=0.6\columnwidth]{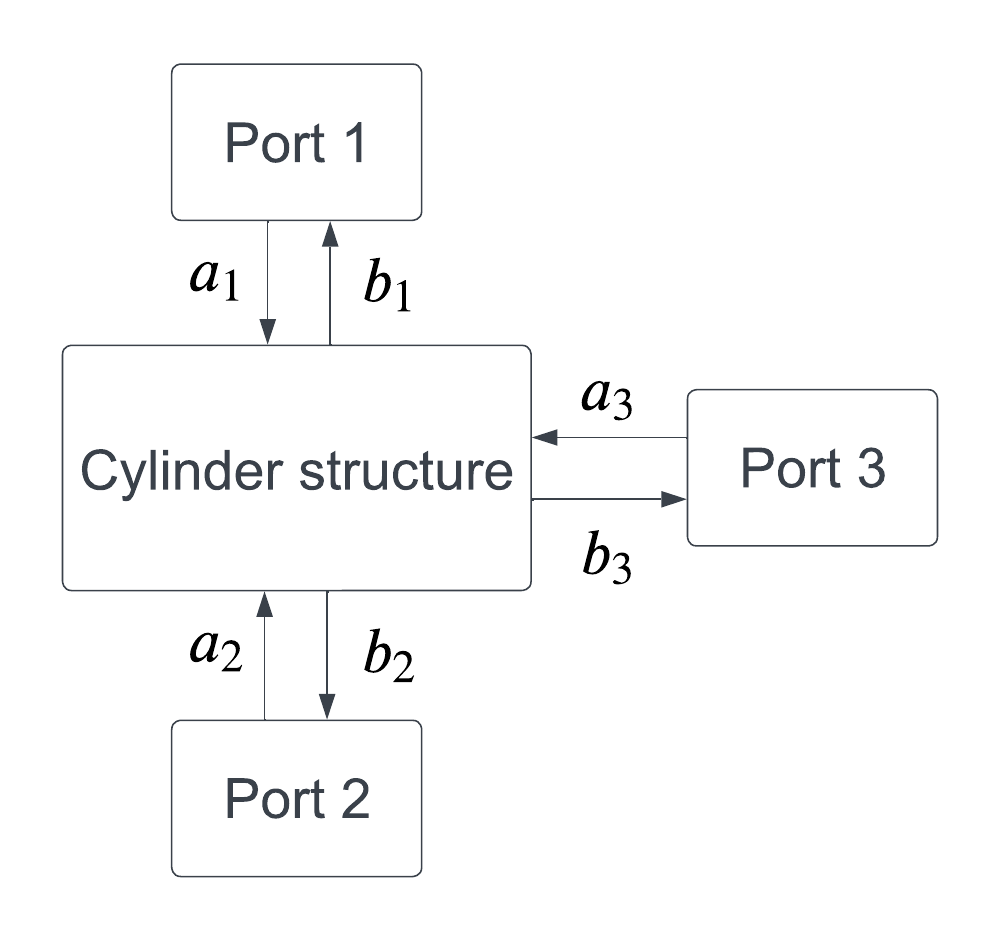}
    \caption{Flow-chart of the three-port electrical network of the LiView system and cylinder. Port 1 capacitively excites the signal, whose reflection can then be detected by port 2. Port 3 corresponds to the cylinder tube and the piston.}
    \label{fig:three_port_model}
\end{figure}

In the present frequency regime, it is common to employ a formalism based on wave quantities that describe scattering on the linear electrical network that represents the investigated system \citep{white2004high}, \citep{ludwig2000}. Given the incident and reflected voltage wave, $E_{i}^{inc}$ and $E_{i}^{ref}$, of a port $i$, we define two waves $a_i$ and $b_i$ as
\begin{align}
	a_i&=\frac{E_{i}^{inc}}{\sqrt{Z_0}},\\ 
	b_i&=\frac{E_{i}^{ref}}{\sqrt{Z_0}},
\end{align}
where $Z_0$ is the transmission line's characteristic impedance, which connects the $i^{th}$ port. With the help of $a$ and $b$ waves, a linear electrical network can be characterized by a set of coupled algebraic equations describing the reflected waves from each port in terms of the incident waves at all the ports \citep{white2004high}. The coefficients of the network are called the scattering parameters $S$. For a three-port electrical system as the LiView system, the network can be described in the matrix format:
\begin{equation}
    \begin{pmatrix}
    b_1\\
    b_2\\
    b_3
    \end{pmatrix}
    =
    \begin{bmatrix}
    S_{11} & S_{12} & S_{13}\\
    S_{21} & S_{22} & S_{23}\\
    S_{31} & S_{32} & S_{33}\\
    \end{bmatrix}
    \begin{pmatrix}
    a_1\\
    a_2\\
    a_3
    \end{pmatrix}.
    \label{eq:matrix_three_port}
\end{equation}
\autoref{fig:three_port_model} provides an abstract illustration of the three-port network which models the LiView system. It is a linear electrical network that interacts with three ports. Port 1 and 2 correspond to probe 1 and 2 in \autoref{fig:liview} respectively. They are connected to the LiView electronic unit via coaxial wires. When measuring, a stimulus is injected at Port 1, and its reflection is finally received by Port 2. Port 3 represents the cylinder tube with the piston moving inside, whose signal can't be measured by any probe. The cylinder structure among the three ports can be considered as the virtual transmission lines. In one measuring cycle, The LiView electronic unit first generates a signal at a predefined frequency and injects it into the cylinder structure through port 1. The injected signal passes from the cylinder bearing to the piston rod which is represented by port 3. It then propagates as electromagnetic waves through the oil inside the cylinder tube until it reaches the piston. The piston then reflects the signal, and the probe at port 2 can then detect this reflected signal. 

The quantities that are measured by the LiView electronic unit are the transmission $t$ (at one specific frequency) between the exciting port 1 and receiving port 2 given by 

\begin{equation}
    t = \frac{b_2}{a_1}.
    \label{eq:def_t}
\end{equation}

The piston position $L$, which is formally defined as the distance between the piston and its \acrfull{bdc} along the axial axis of the cylinder, determines the scattering parameters of the 3rd port:
\begin{equation}
    a_3 = T(L)b_3,
    \label{eq:scat_params_third_port}
\end{equation}
with $T(L)$ modeling the cylinder as a transmission line. Under the assumption of reciprocity $S_{ij} = S_{ji}$, one can extract the relevant transmissions and reflections at ports 1 and 2 by combining \mautoref{eq:matrix_three_port,eq:scat_params_third_port}. Exemplary, the transmission $t$ for a certain frequency as in \autoref{eq:def_t} for low frequency TEM-modes then reads \citep{Scheidt2017}
\begin{equation}
    t = S_{21} - \frac{S_{23}S_{13}}{S_{33} - \frac{1}{T(L)}}.
    \label{eq:transmission}
\end{equation}
Note that all wave-related variables ($E, a, b, S, t$) in previous equations are complex numbers. With Euler's formula, it follows that any complex number $z$, with $ r=\vert z\vert$ as its amplitude and $ \theta={\mathrm arg}(z)$ as its phase, can be written as
\begin{align}
    \begin{split}
        z &= r {\mathrm e}^{ {\mathrm j} \theta} \\
            &=r \cos\theta + {\mathrm j} r \sin\theta \\
            &={\mathrm Re}(z)+{\mathrm j} {\mathrm Im}(z)  
    \end{split}
    \label{eq:euler}
\end{align}
where ${\mathrm j}=\sqrt{-1}$ so that the real and imaginary part of $z$, ${\mathrm Re}(z)$ and ${\mathrm Im}(z)$, are both real numbers. Therefore, the data finally gathered by the LiView system are a complex-valued vector $\bm{t}$ which has 121 elements corresponding to transmissions at the 121 frequencies, equally distributed in a range from \SI{300}{\mega\hertz} to \SI{1.5}{\giga\hertz}.

Our physical model for position detection is based on the electrodynamics of the cylinder tube given by \autoref{eq:transmission} and the derivation of $T(L)$  from Maxwell's Equations according to \cite{Scheidt2017}. Specifically, 
\begin{equation}
    T(L) = {\mathrm e}^{2L(-\epsilon^{\prime\prime}+\mathrm{j}k}),
    \label{eq:t_of_l}
\end{equation}
where $\epsilon''$ represents the imaginary part of the complex-valued relative permittivity $\epsilon_r = \epsilon'+j\epsilon''$ and $k$ refers to the angular wavenumber. 

The existence of the relation $t=f(L)$ (an abstraction of \mautoref{eq:transmission,eq:t_of_l}) paves the way towards a position detection method by solving the inverse $L=f^{-1}(\bm{t})$. Our physical model directly gives out the piston position by solving the inverse function. Looking at \mautoref{eq:transmission,eq:t_of_l}, it's apparent that piston position $L$ is nothing but a mathematical expression of the transmission $\bm{t}$, which can be measured by the LiView system, subjecting to scattering parameters of the three-port electrical network (\autoref{fig:three_port_model}), the relative permittivity of the oil inside the cylinder, and the angular wavenumber. We assume all variables except the model input $t$ and output $L$ are constants during inference and determine their values in advance. The scattering parameters $S_{21}$, $S_{33}$, as well as the product of $S_{23}$ and $S_{13}$ in \autoref{eq:transmission} are dependent on the specific cylinder structure and acquired by calibration via measurements or simulations of the respective cylinder. The relative permittivity $\epsilon_r$, which quantifies the storage (real part) and dissipation (imaginary part) of the energy, can be obtained with the help of a Vector Network Analyzer (VNA). Since we know the frequency for each measurement cycle of the LiView electronic unit, we can easily get the angular wavenumber $k$ from the equation $k=2 \pi f \sqrt{\epsilon'\mu_r}/c_0$ where $f$ is the frequency, $\epsilon'$ refers to the real part of the relative permittivity, $\mu_r$ stands for the relative permeability which is approximately equal to 1 for oil and $c_0$ is the speed of light in vacuum. After all the calibrations, we could finally substitute the predetermined variables and the transmission $t$ measured by the LiView system during inference to solve the desired piston position $L$

As long as the predetermined variables keep unchanged and the modeling of the coaxial cylinder structure remains accurate, the physical model performs well both in theory and in practice. However, the assumptions can surely break under harsh 
and irregular environments. The oil permittivity $\epsilon$ is sensitive to environmental factors such as temperature, impurities in the oil, etc. The temperature effect is even more severe at extremely high or low temperatures. Because then, the change in permittivity of other materials such as the seals could break the assumption of the constant scattering parameters. To solve this issue within the domain of the physical method, we have to implement a mechanism for the physical model to decide when to use which calibrated scattering parameters. This method is however inaccurate, complicated to implement, and leads to noisy position values. Moreover, the coaxial structure is only a good approximation if the propagation length of the signal is long. As a result, when the piston is near the bottom of the cylinder tube or the piston rod bearing, the model assumptions of a coaxial cavity do not hold anymore and workarounds such as a switch to other models or look-up tables have to be implemented, which demand in-depth expert knowledge and impede a robust implementation that can easily be generalized to mechanically differing cylinder types. Then more elaborate and general models are needed for accurate position determination. Next, we will show how we found  these more accurate models with better generalization performance by switching to learning-based methods. 

\section{Methods}
\label{sec:methods}

We apply machine learning and deep learning-based approaches as a generic solution to cylinder piston position estimation by learning the underlying relationship between the input measurement and the target piston position, with the help of a large amount of data collected in a variety of work scenarios. Specifically, our goal is to learn the mapping relationship $L=f^{-1}(\bm{t})$ between the transmission $\bm{t}$ measured by the LiView system and the desired piston position $L$ by machine learning and deep learning methods. The original measurement data are 1D complex-valued vectors with 121 elements corresponding to the 121 equally distributed frequencies from \SI{300}{\mega\hertz} to \SI{1.5}{\giga\hertz}. Depending on the model architecture and the technology applied, we preprocess the raw measurement into the actual input to the model accordingly (sections \ref{sec:dl}, \ref{sec:cvnn}, and \ref{sec:fe}). The output of all models is always a scalar referring to the piston position recorded simultaneously when the input is measured.

This section introduces the general architectures of the models and specific methods we use to achieve our goal. we apply both the classical machine learning (section \ref{sec:ml}) and deep learning-based models (section \ref{sec:dl} and \ref{sec:cvnn}), together with a novel technique named \acrlong{fe} (\ref{sec:fe}). The implementation details are given in section \ref{sec:implementation} and the model performance is analyzed in section \ref{sec:comparison}.

\subsection{Machine Learning Models}
\label{sec:ml}

\textbf{\acrfull{rf}.} \acrlong{rf} is a combination of tree predictors such that each tree depends on the values of a random vector sampled independently and with the same distribution for all trees in the forest \citep{breimanRandomForests2001}. To construct a \acrshort{rf}, we first generate $m$ new training sets from the original training set of size $N$ and feature dimension $d$ (with $d=242=121\times2$ since we use real-valued model until section \ref{sec:cvnn}) by uniform sampling with replacement. This technique is also known as bootstrapping. For each sampled new training set, we randomly sample $d^{\prime}=\lfloor\sqrt{d}\rceil$ features and then train \acrshort{cart} \citep{reason:BreFriOlsSto84a} on the sampled feature subspace. Finally, all the trained \acrshort{cart} are aggregated to form the \acrshort{rf}, and the average prediction result of the trees is the final output. The process of bootstrapping and then aggregating is often abbreviated as bagging.

\textbf{\acrfull{gbdt}.} Another classical machine learning method based on the aggregation of decision trees is \acrshort{gbdt}. The main difference between \acrshort{gbdt} and \acrshort{rf} is that the former uses boosting instead of bagging as in the latter. Boosting is a powerful technique for combining multiple 'base' classifiers to produce a form of a committee whose performance can be significantly better than any of the base classifiers \citep{bishop2006pattern}. In our case, we use \acrshort{dt}s as base regressors and build an ensemble of them in a forward stagewise manner. At each iteration of the forward stagewise procedure, we train a \acrshort{dt} to represent the negative gradient of the residual error from the previous iteration and aggregate to the ensemble by steepest descent.

\acrshort{rf} and \acrshort{gbdt} are both ensemble learner based on \acrshort{dt}s. They utilize bagging and boosting respectively to overcome the tendency of overfitting and high variance in a single \acrshort{dt} for better generalization performance. Given the excellent performance of both \acrshort{gbdt} \citep{chenXGBoostScalableTree2016,dorogushCatBoostGradientBoosting2018,keLightGBMHighlyEfficient2017} and \acrshort{rf} \citep{breimanRandomForests2001,schonlauRandomForestAlgorithm2020} on general regression problems, we adopted these two models for our task. However, these two models do not perform very well on our problem (section \ref{sec:comparison}), especially compared to the deep learning-based models presented in the next section.

\subsection{Deep Learning Models}
\label{sec:dl}

\begin{figure}[h]
	\centering
	\begin{subfigure}[b]{0.4\columnwidth}
		\centering
		\includegraphics[width=0.8\textwidth]{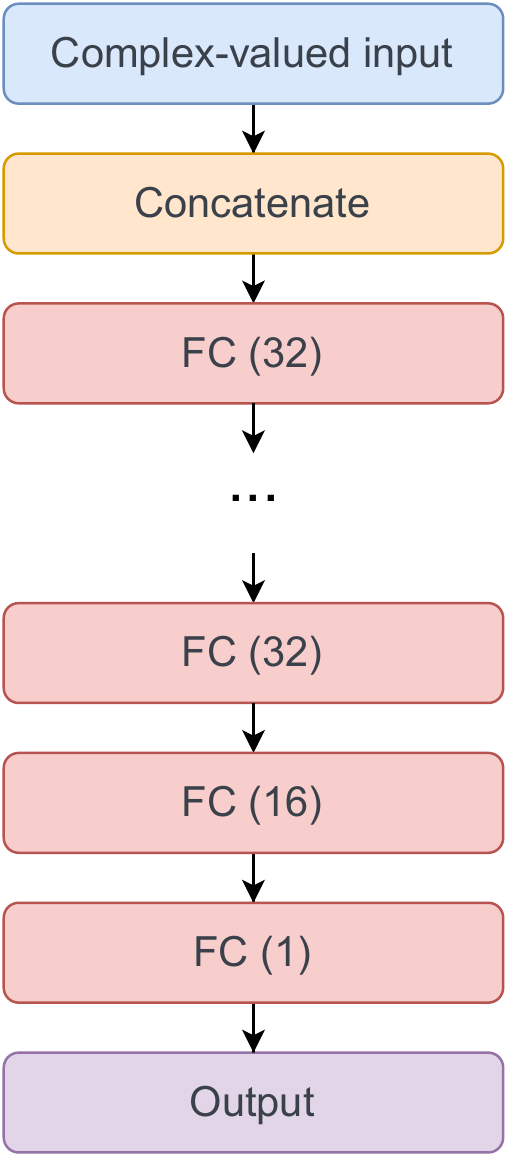}
		\caption{Real-valued MLP}
		\label{fig:rvmlp}
	\end{subfigure}%
% 	\hfill
	\begin{subfigure}[b]{0.4\columnwidth}
		\centering
		\includegraphics[width=0.8\textwidth]{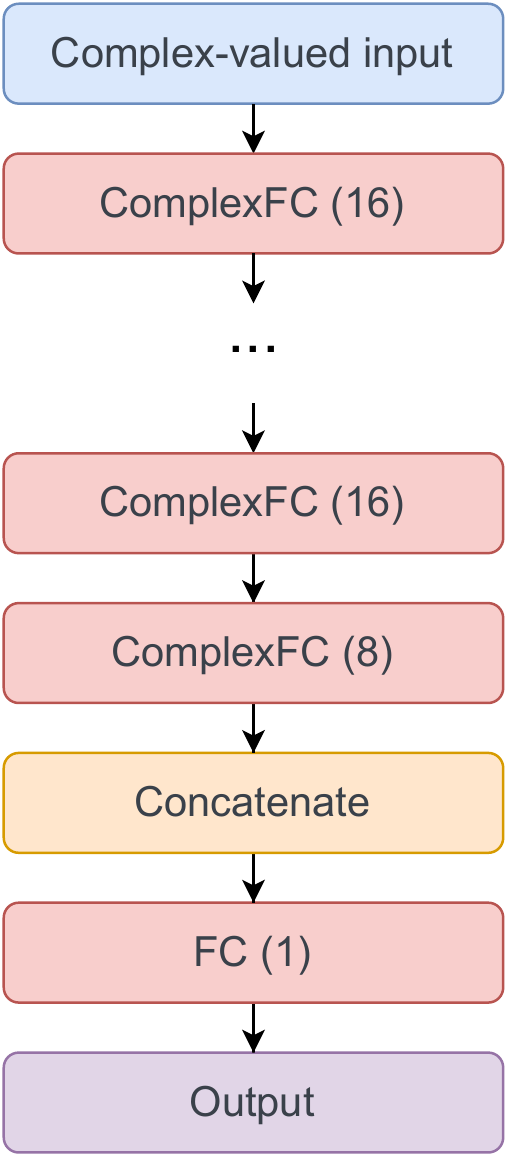}
		\caption{Complex-valued MLP}
		\label{fig:cvmlp}
	\end{subfigure}%
	\caption{MLP architecture. The real-valued (a) and complex-valued (b) \acrshort{mlp}s consist of Fully Connected (FC) and Complex-valued Fully Connected (ComplexFC) layers, respectively, with the number of units of a specific layer given in its following parentheses.}
	\label{fig:mlp}
\end{figure}

\begin{figure}[] 
	\centering
	\begin{subfigure}[b]{0.4\columnwidth}
		\centering
		\includegraphics[width=0.8\textwidth]{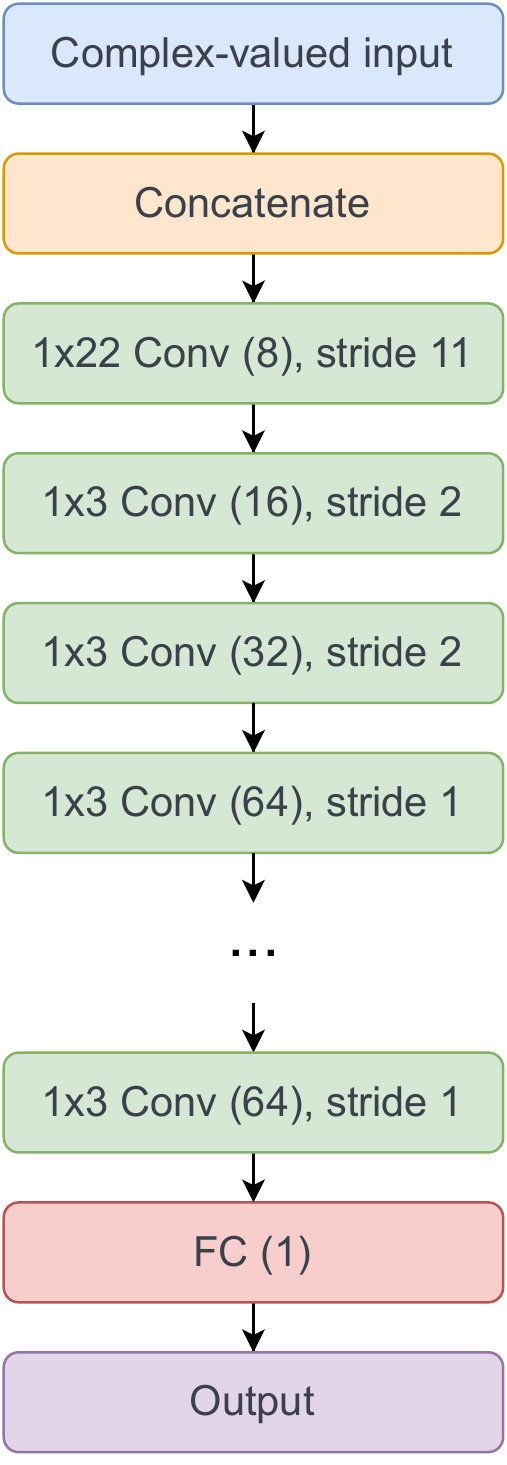}
		\caption{Real-valued CNN}
		\label{fig:rvcnn}
	\end{subfigure}%
% 	\hfill
	\begin{subfigure}[b]{0.5\columnwidth}
		\centering
		\includegraphics[width=0.8\textwidth]{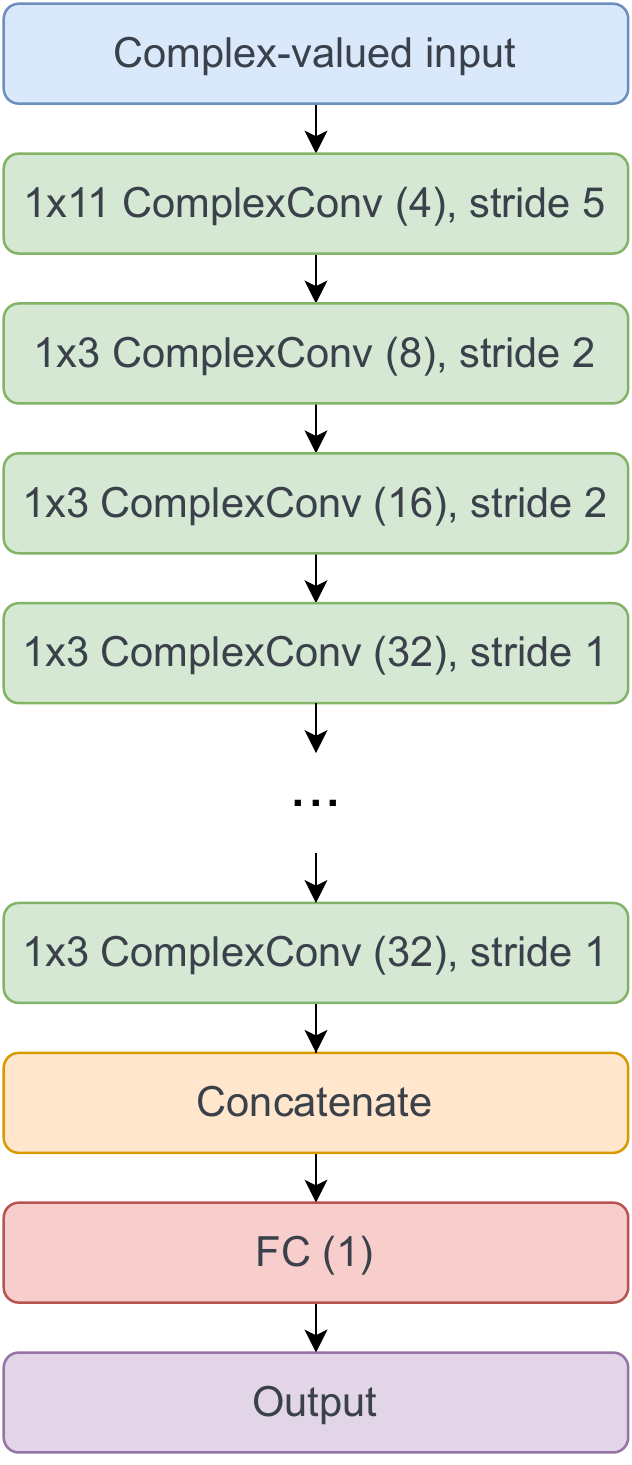}
		\caption{Complex-valued CNN}
		\label{fig:cvcnn}
	\end{subfigure}%
	\caption{CNN architecture. The backbone of the real-valued (a) and complex-valued (b) \acrshort{cnn}s is 1D convolutional (Conv) and 1D complex-valued convolutional (ComplexConv) layers, respectively. For each layer in the illustration, the kernel size is presented as the first term of the description text and the number of output channels is given in parentheses. Different strides are used for downsampling. At the end of all \acrshort{cnn}s, there is always a Fully Connected (FC) layer to map the features extracted by Conv or ComplexConv layers to the desired output. In practice, we modify the number of convolutional layers so that each model has around 10K parameters.}
	\label{fig:cnn}
\end{figure}

Deep learning, or \acrshort{ann}, allows computational models composed of multiple processing layers to learn data representations with multiple levels of abstraction \cite{lecunDeepLearning2015}. Intricate patterns in large data sets can be implicitly discovered by exploiting backpropagation \citep{rumelhartLearningRepresentationsBackpropagating1986} to change its internal parameter states. The motivation for the choice of deep learning is to use this mechanism to achieve better generalization performance since traditional machine learning methods are insufficient to learn complicated functions in high-dimensional spaces such as our case (with evidence given in section \ref{sec:comparison}). The architecture of a deep learning model is a multilayer stack of simple modules (\mautoref{fig:mlp,fig:cnn,fig:cvnn}). In this section, we discuss the structure of our \acrshort{mlp}, \acrshort{cnn},  and \acrshort{cvnn} models.

\textbf{\acrfull{mlp}.}
\label{sec:mlp}
\acrshort{mlp} \citep{werbosRegressionNewTools1974,rumelhartLearningRepresentationsBackpropagating1986} is a feedforward \acrshort{ann} where every single neuron in adjacent layers is connected to each other. Because of the fully connected characteristic, \acrshort{mlp} is also named as \acrfull{fcnn}. The quintessential advantage of \acrshort{mlp}s is that they are universal approximates provided sufficiently many hidden units \citep{hornikMultilayerFeedforwardNetworks1989}, i.e. they are able to represent a wide variety of functions requiring no special assumptions about the input. For the real-valued \acrshort{mlp} (\ref{fig:rvmlp}), we first concatenate the real and imaginary parts of a measured complex-valued transmission $\bm{t}$ to form a real-valued vector as the new input, and train real-valued \acrshort{mlp}s to approximate the inverse function $L=f^{-1}(\bm{t})$ for prediction of the targeted piston position $L$. We evaluate \acrshort{mlp}s with different numbers of hidden layers, each consisting of 32 neurons except the last one, which has 16 neurons. We also analyze the effect of four different activation functions, sigmoid, ReLU \citep{nairRectifiedLinearUnits}, Leaky ReLU \citep{maasRectifierNonlinearitiesImprove}, and SELU \citep{gunterklambauerSelfNormalizingNeuralNetworks2017}, on \acrshort{mlp}s. 

\textbf{\acrfull{cnn}.} \acrshort{cnn}s \citep{lecunBackpropagationAppliedHandwritten1989} are a specialized kind of neural network for processing data that has a known, grid-like topology \citep{Goodfellow-et-al-2016}. Units in a convolutional layer are organized in feature maps, within which each unit is connected to local patches in the feature maps of the previous layer through a set of weights \citep{lecunDeepLearning2015}. The result of this locally weighted sum is then passed through the activation function like in \acrshort{mlp}s to form the output feature map. 

The motivation for \acrshort{cnn} is its property that a convolutional layer should learn functions that represent local interactions in input. In our case, the local interactions are among neighboring frequencies. Based on the properties of electromagnetic waves, it is reasonable to assume that transmission $t$ of adjacent frequencies are highly correlated, exhibiting local statistics which are easily detected. In section \ref{sec:comparison}, we prove our hypothesis with training results and provide a credible explanation for the underlying physical mechanism.

As in real-valued \acrshort{mlp}, we concatenated the real-valued real and imaginary parts of all 121 complex numbers to form the 1D grid with 242 elements as the input for our real-valued \acrshort{cnn}s. Once we converted the complex-valued input to vectors with the shape of $1\times242$, 1D local convolutional filters only to the frequency dimension (instead of the temporal dimension). A 1D \acrshort{cnn} can then be constructed using these convolutional layers and trained to predict the targeted piston position. As depicted in \autoref{fig:rvcnn}, the first convolutional layer has a large kernel size ($1\times22$) and stride (11). It's a common way to reduce dimensionality in frequency-based signal processing, e.g. in \cite{supriyaTriggerWordRecognition2020}. The first convolutional layer is followed by three convolutional layers with small kernels ($1\times3$) and strides to extract informative features. Finally, a fully connected layer exploits those extracted features to make the prediction. 

Note that we perform downsampling directly by convolutional layers, instead of max pooling layers, with a stride of more than one. The two ways result in a similar model structure considering the dimension of the feature maps of each layer and are experimentally proved in section \ref{sec:comparison} to have similar effects on generalization performance for our project. We prefer the latter from the perspective of intuitive comparisons between real-valued and complex-valued \acrshort{cnn}, described in the next section.

\subsection{\acrfull{cvnn}}
\label{sec:cvnn}

\begin{figure}
    \centering
    \includegraphics[width=0.9\columnwidth]{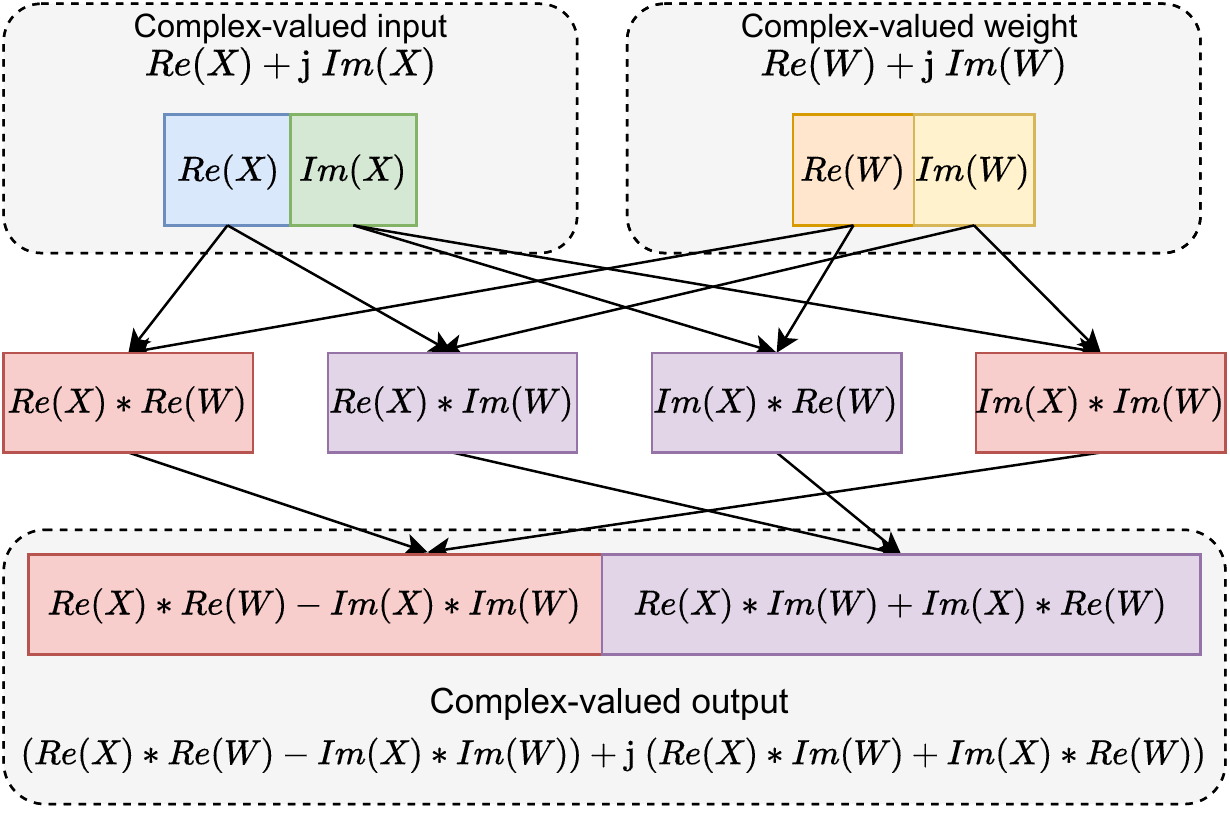}
    \caption{An illustration of one layer in a \acrshort{cvnn}. Both the input and the output are complex numbers. Layers in\acrshort{cvnn} mimic calculations in the complex domain. The symbol $*$ in the figure represents multiplication for ComplexFC layer and convolution for ComplexConv layer.}
    \label{fig:cvnn}
\end{figure}

\acrshort{cvnn}s are \acrshort{ann}s that operate in complex space with complex-valued input and weights. The main reason for their advocacy lies in the difference between the representation of the arithmetic of complex numbers, especially the multiplication operation \citep{basseySurveyComplexValuedNeural2021}. \autoref{fig:cvnn} illustrates our implementation of the complex-valued operation inside a layer of a \acrshort{cvnn}. In the illustration, $Re(X)$ and $Im(X)$ refer to the real and imaginary parts of a complex-valued input, while $Re(W)$ and $Im(W)$ refer to the real and imaginary parts of the layer weight. To imitate the multiplication of complex numbers, we first split the real and imaginary parts of the complex-valued input and weight as stated, then calculate the components $Re(X)*Re(W)$, $Re(X)*Im(W)$, $Im(X)*Re(W)$, and $Im(X)*Im(W)$ separately, and lastly additively combine these four products to construct a new complex-valued output of the specific layer with the real part as $(Re(X)*Re(W)-Im(X)*Im(W))$ and the imaginary part as $(Re(X)*Im(W)+Im(X)*Re(W))$. Once the complex-valued outputs are obtained, a split activation function \citep{basseySurveyComplexValuedNeural2021} is applied to them, where the real and imaginary components of the signal are independently input into a chosen real-valued activation function.

Note that the symbol $*$ here does not have to refer to the multiplication operation. One can easily switch the multiplication operation to the convolution operation. In this case, $X$ will refer to complex-valued feature maps from the last layer and $W$ the complex-valued kernels of a convolutional layer. Accordingly, $Re(X)$ and $Im(X)$ (or $Re(W)$ and $Im(W)$) each represent half of the feature maps (or kernels). A great illustration of the operation of a complex-valued convolutional layer can be found in \cite{trabelsiDeepComplexNetworks2018}. Furthermore, we can extend the meaning of $*$ to an arbitrary function evaluation. In this case, $Re(W)$ and $Im(W)$ are the functions in the same form but with different coefficients. They take either $Re(X)$ or $Im(X)$ as input and the outputs are denoted as the four blocks in the second row of \autoref{fig:cvnn}.

Inspired by \cite{PhysRevX.11.021060} and \cite{trabelsiDeepComplexNetworks2018}, we implement the $Re(W)$ and $Im(W)$ as two standard real-valued layers respectively but combine their outputs in the way stated above to construct a \acrshort{cvnn} instead. If we denote the  We denote the implementation of complex-valued counterparts of FC() and Conv() as ComplexFC() and ComplexConv() (in \mautoref{fig:mlp,fig:cnn}), with the number of neurons for FC layer or kernels for Conv layer in their real-valued layers given in the following parentheses. For example, a ComplexFC(16) layer is implemented by two FC(16) layers representing $Re(W)$ and $Im(W)$, respectively.

The choice of \acrshort{cvnn} is motivated by the fact that the input is complex numbers, where the real and imaginary parts, or more specifically their amplitude and phase, are statistically correlated. This information will be lost if we simply choose to represent a complex number as an ordered pair of two real numbers, as in the previous sections. \acrshort{cvnn} can reduce the ineffective degree of freedom in learning or self-organization to achieve better generalization characteristics \citep{hiroseComplexValuedNeuralNetworks2012}. The mechanism of \acrshort{cvnn} can also be seen as applying an infinitely strong prior to the original \acrshort{mlp} and \acrshort{cnn} forcing them to operate in the complex domain. This could be (and is proved in section \ref{sec:comparison} to be) useful since we're giving the models more information about the real-world application. Guided by this idea, we further introduce another technique named \acrlong{fe}.

\subsection{\acrfull{fe}}
\label{sec:fe}

\begin{figure}
    \centering
    \includegraphics[width=0.8\columnwidth]{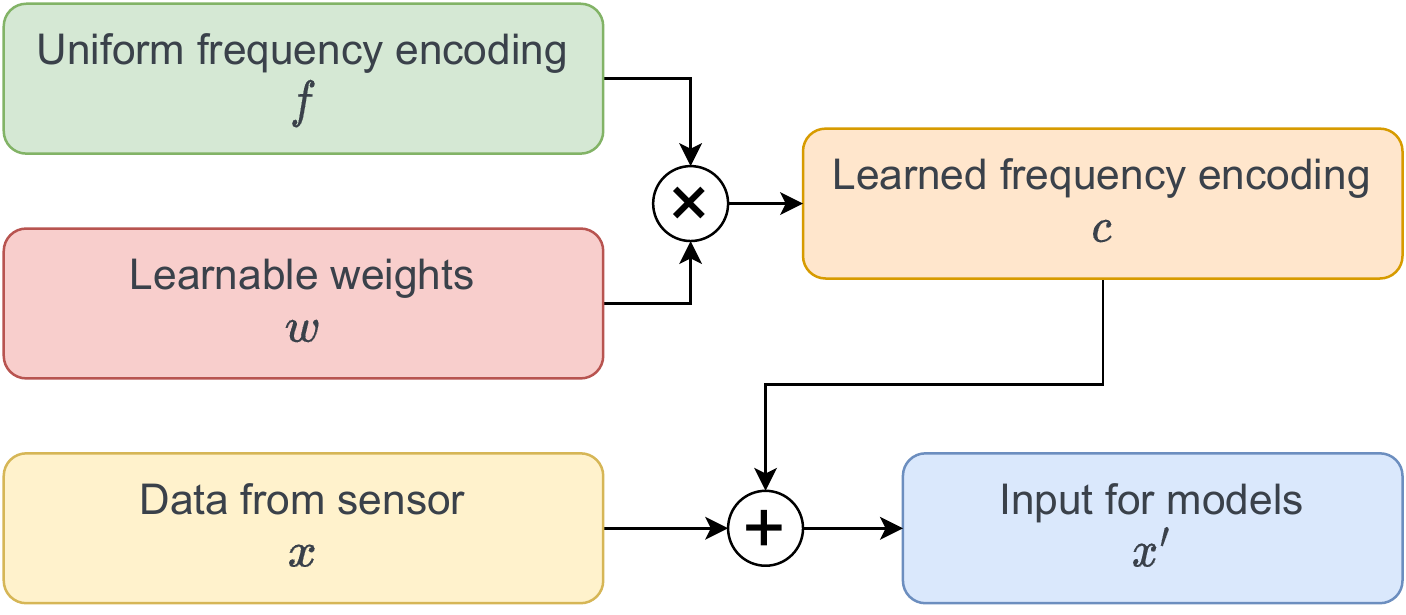}
    \caption{Sketch of the \acrfull{fe} procedure. The normalized frequencies $\bm{f}$ are multiplied with the learnable weights $\bm{w}$ which leads to the code $\bm{c}$. This $\bm{c}$ then serves as an additive bias for the original model input $\bm{x}$. Therefore, the frequency information is encoded and injected to the final input for models $\bm{x^{\prime}}$ }
    \label{fig:fe}
\end{figure}

\acrshort{fe} encodes the (normalized) frequencies $\bm{f}$ for each channel of the complex-valued input, and then the code $\bm{c}$ is added element-wise to the input after weighting by a learnable parameter $\bm{w}$. Formally, the \acrshort{fe} pipeline in this paper can be expressed as:
\begin{equation}
    \begin{split}
        \bm{c} &= \bm{w}*\bm{f},\\
        Re(\bm{x^{\prime}}) &= Re(\bm{x}) + \bm{c},\\
        Im(\bm{x^{\prime}}) &= Im(\bm{x}) + \bm{c},
    \end{split}
    \label{eq:fe}
\end{equation}
where $\bm{x^{\prime}}$ and $\bm{x}$ are the encoded and original 1D input vector, respectively. Both consist of 121 elements corresponding to the 121 frequencies. As illustrated in \autoref{fig:fe}, the 121 frequencies from \SI{300}{\mega\hertz} to \SI{1.5}{\giga\hertz} are uniformly scaled to [0,1] at first. The normalized frequency vector is denoted as $\bm{f}$. Then the uniformly encoded frequencies $\bm{f}$ are multiplied by the corresponding learnable weights to generate the code $\bm{c}$. Note that the weights $\bm{w}$ for frequency encoding are learned by backpropagation during training, just like any other learnable parameters in the model. After that, the code for each frequency is added to both the real and imaginary parts of the original input $\bm{x}$ (in our case the measured transmission $t$) at the corresponding frequency, forming the new input under \acrshort{fe} $\bm{x^{\prime}}$ to the models. 

\acrshort{fe} injects the information about the exact frequency of the electromagnetic waves that the sensor operates on directly into the model. Frequency information should (and in section \ref{sec:comparison} proved to) be helpful because the precise value of the frequency will significantly affect the corresponding transmission $t$, and furthermore, the piston position. Since we cannot precisely confirm a priori the impact of each frequency, the learnable weights are introduced to correct the initial uniform encoding. We prefer \acrshort{fe} over intuitive concatenation, mainly for the following reasons. Firstly, it uses the summation of corresponding elements to avoid the explosion of the input dimension and further the model parameters. Secondly, the frequency information is accurately introduced into the correct corresponding input element instead of letting the model find the correspondence between the two by training itself. Another advantage of \acrshort{fe} is that it's also a model-agnostic technique, i.e., we can easily apply it to any learning-based models benefiting from backpropagation without any particular change of the model structure. It turns out that \acrshort{fe} significantly affects model performance. More details about the evaluation results can be found in section \ref{sec:comparison}. 

\section{Experiments}
\label{sec:experiments}

\subsection{Dataset}
\label{sec:dataset}
\begin{figure}
    \centering
    \includegraphics[width=\columnwidth]{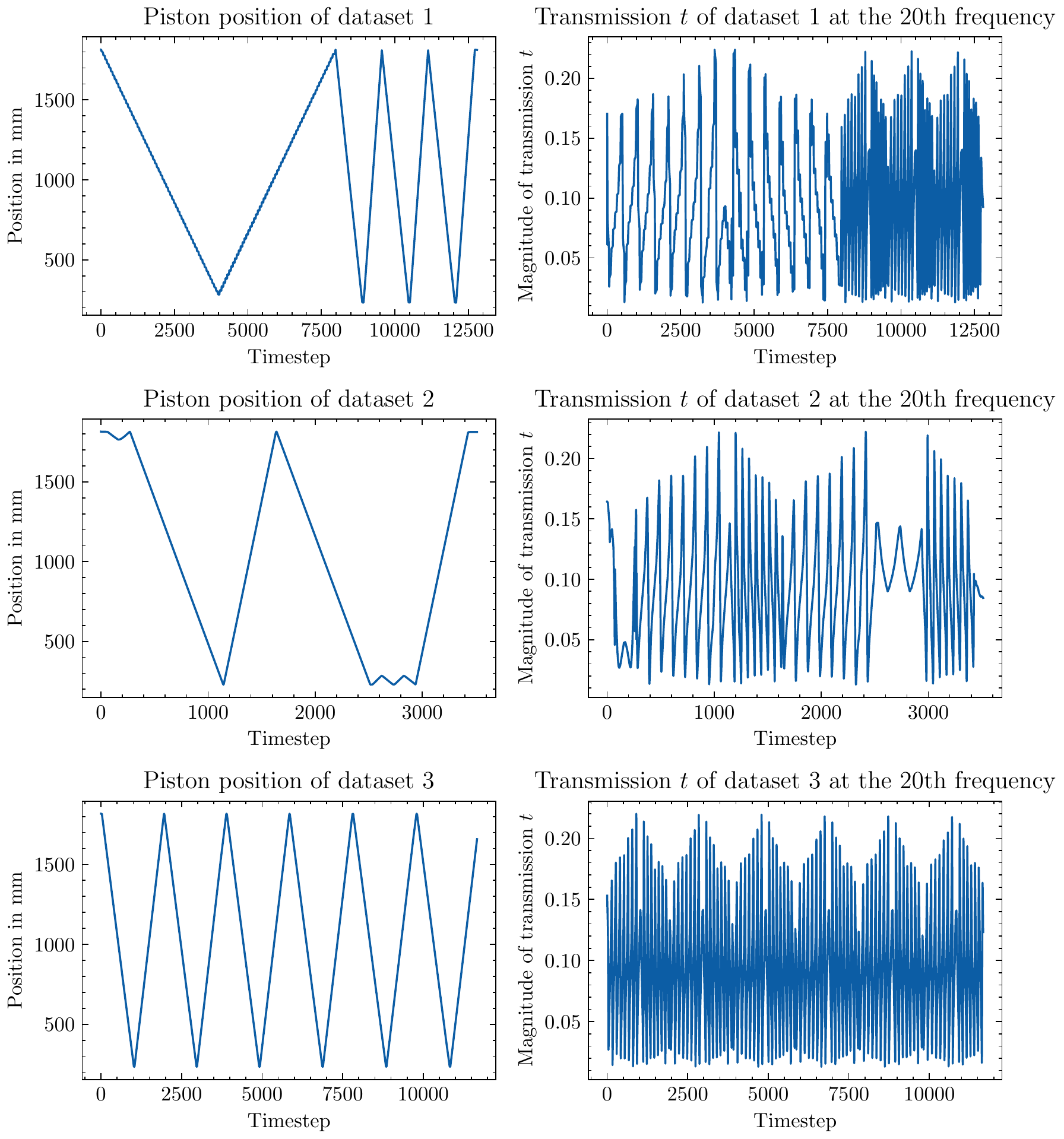}
    \caption{Position and S21 parameters from three typical datasets. Each row represents one dataset, with its piston position profile showing on the left and the complex magnitude (or amplitude) of the measured transmission at the 20th frequency on the right. }
    \label{fig:datasets}
\end{figure}

\textbf{Experimental Configuration.} Our experiments for the measurement of the transmission $\bm{t}$ and the recording of the piston position $L$ are carried out at a wide variety of temperatures in the range from \SIrange{25}{95}{\celsius}. In all experiments, a LiView device is mounted on the same hydraulic cylinder with a stroke of \SI{1815}{\milli\meter} and keeps measuring transmissions at 121 frequencies uniformly distributed in a range from \SI{300}{\mega\hertz} to \SI{1.5}{\giga\hertz} as described in section \ref{sec:liview}. A hydraulic power unit drives the cylinder, and a magnetostrictive position measuring system measures the piston position. Other than the input and output for the models, metadata such as operating frequency, cylinder temperature, and timestamps are also recorded. Three typical experiment recordings are shown in \autoref{fig:datasets}. As we can see from the three graphs on the left side of \autoref{fig:datasets}, typical piston movements are uniform linear motions shown as straight lines in the figures. To test the performance under different working conditions, we also adopt varying speed profiles by mixing end-to-end and jittering movements at different speeds during different experiments, shown respectively in different slopes of lines in the upper left and the zigzag patterns in the middle. The amplitudes of the corresponding measured input are drawn on the right side of \autoref{fig:datasets}. It's evident that the transmission $\bm{t}$ changes with the piston position $L$, which verifies our assertion of the existence of $t=f(L)$.

\begin{figure}
    \centering
    \includegraphics[width=0.8\columnwidth]{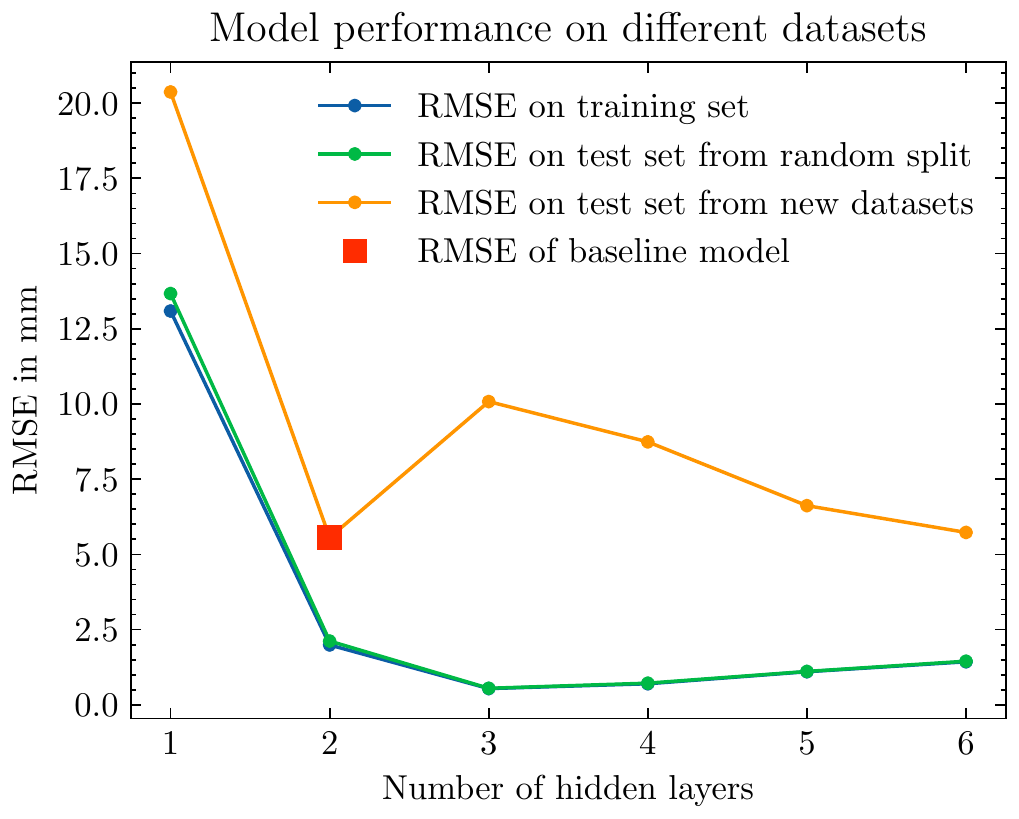}
    \caption{Performance of \acrshort{mlp} models with different numbers of hidden layers on different datasets}
    \label{fig:rmse_vs_hidden}
\end{figure}

\textbf{Dataset Split.} The core idea behind our dataset split is that we selected two test sets deliberately in two different ways to test which one reflects the true generalization ability of our models. We generated 534K data pairs $(\bm{t}, L)$ among 73 experiments in total, where $\bm{t}$ is a 1D vector with 121 complex-valued elements corresponding to transmissions at 121 frequencies and $L$ stands for the targeted piston position. We randomly split the data from 68 datasets at a ratio of 80 to 20 for the training and the first test set. There is also a second test set composed of the remaining five datasets. To distinguish the two test sets, we named them "test set from random split" and "test set from new datasets" (since the latter one is "new" to the training set) in \autoref{fig:rmse_vs_hidden}, respectively. The final ratio of the data volume in three datasets is 10/2/1 (training/test1/test2).

We split the data this way to show that only the second test set, "test set from new datasets", is the only suitable set to test generalization. As can be seen from \mautoref{fig:rmse_vs_hidden}, the performance of all \acrshort{mlp} models on the first test set, "test set from random split", is almost identical to the performance on the corresponding training sets. We argue that, although both test sets are kept unseen from the training process, the "test set from the random split" shares too many influential factors with the training set since they are from the same experiments, which makes the data in the two not quite the same but very close. These influential factors could be explicit ones such as speed profile and temperature, or implicit ones like chamber pressure, oil quality, parts aging effect, etc. The second test set, consisting of complementary datasets, doesn't suffer from the issue and therefore is the only suitable test set that reflects the actual generalization performance. That said, the test set from the random split is not entirely useless. It is still an indicator to show what will happen if there are only slight changes, which was helpful in the model selection phase and could help if we could cover a broader range of working conditions of the cylinder. But in the following sections, unless otherwise specified, we refer to "the test set" as the one from the new datasets.

\subsection{Implementation Details}
\label{sec:implementation}

This section describes the implementation of the models introduced in \mautoref{sec:methods} and the training details.

Since the input for our models are 1D vectors with 121 complex-valued elements, we treat them for both classical machine learning models and real-valued \acrshort{ann}s as real-valued input with 242 features by concatenating their real and imaginary parts. On the other hand, \acrshort{cvnn}s can use the measured data with 121 complex-valued features off-the-shelf. We also tried to represent complex numbers using magnitude and phase, but the final results of the two ways to handle complex numbers were almost indistinguishable. \acrlong{fe} was also applied to the models as stated in section \ref{sec:methods} by choice. 

To preserve the possibility of running the model on an edge computing device and integrating it into the current LiView electronic unit in the future, we limited the total parameters for each model to just under 10K.

We utilized the scikit-learn \citep{pedregosa2011scikit} implementation of random forest regressor. For \acrshort{gbdt}, we used LightGBM \citep{keLightGBMHighlyEfficient2017}, which can in theory significantly outperform XGBoost \citep{DBLP:journals/corr/ChenG16}, another popular implementation of \acrshort{gbdt}, in terms of computational speed and memory consumption according to \cite{keLightGBMHighlyEfficient2017}. For both \acrshort{rf} and \acrshort{gbdt}, we fine-tuned the hyperparameters in 500 trials sampled by TPE (Tree-structured Parzen Estimator) algorithm in a large hyperparameter space which covered both shallow and deep models with the help of the hyperparameter optimization framework Optuna \citep{akibaOptunaNextgenerationHyperparameter2019}. The test performance of both models with the best hyperparameter set is given in \autoref{tab:ml}.

We also trained all \acrshort{ann}s, implemented by PyTorch 1.12 \citep{NEURIPS2019_9015}, using the hyperparameters acquired by fine-tuning in 500 trials with the help of Optuna. It turned out that the best hyperparameter sets for \acrshort{mlp}, \acrshort{cnn} and \acrshort{cvnn} were very similar. Accordingly, we chose the final hyperparameters as the following. At the beginning of the training process, we always initialized the biases to zeros and the weights using Xavier uniform random initialization procedure \citep{glorotUnderstandingDifficultyTraining2010}. For training, we used the AdamW optimizer \citep{loshchilovDecoupledWeightDecay2017}, which is the combination of the classical Adam \citep{diederikp.kingmaAdamMethodStochastic2017}, with $\beta_1=0.9 $ and $\beta_2=0.999$, and the decoupled weight decay regularization of which we set the coefficient to $10{-4}$. We chose the learning rate of $10{-3}$ and a learning rate scheduler that linearly decreased the learning rate starting at half of the fixed total epochs from $10{-3}$ to 0. This simple learning rate was proved to achieve better generalization compared to no scheduler or delicate ones such as cosine annealed warm restart scheduler \citep{loshchilovSGDRStochasticGradient2017} for all models. The batch size was set to 128 for \acrshort{mlp}s and 64 for \acrshort{cnn}s. We also applied \acrfull{bn} \citep{ioffeBatchNormalizationAccelerating2015} to some models (\mautoref{tab:act,tab:mc}) and early stopping mechanism to all models. Another regularization we tested was Dropout \citep{srivastavaDropoutSimpleWay2014}, but it did not help the generalization in all our trials. We also compared four activation functions (sigmoid, ReLU, Leaky ReLU, and SELU) under the same model structure and went along only with the best-performing one, sigmoid, after the comparison (section \ref{sec:comparison}).

We trained our models on a Ubuntu workstation with NVIDIA GeForce RTX 3090, AMD Ryzen 9 3950X, and 64 GB DDR4 RAM. The evaluations 
(section \ref{sec:comparison}) were conducted on the same workstation. To validate the performance of our model on edge computing devices, we also tested the inference speed on the CPU of a Jetson Xavier NX with the lowest power mode (10 W, 2 cores).

\subsection{Model Comparison}
\label{sec:comparison}

A detailed discussion of the evaluation results and analysis can be found in this section. We use \acrfull{rmse} on the test set as the major criterion for our analysis of model performance. To facilitate the assessment of the model performance by other standards in the industry, we also provide test \acrfull{me} and \acrfull{mae} for some models. \acrshort{rmse}, \acrshort{me} and \acrshort{mae} are all given in millimeter. Accordingly, we calculate the \acrfull{re} as the ratio between \acrshort{rmse} and the stroke of the cylinder \SI{1815}{\milli\meter}. Since the \acrshort{me} of our models are all close to 0, the more commonly used error expression in industry, which is $\textrm{\acrshort{me}} \pm \textrm{standard deviation}$, can be approximated as $\textrm{\acrshort{me}} \pm \textrm{\acrshort{rmse}}$ in our case.

\textbf{Baseline Model.} We started our analysis by setting up the baseline model. In consideration of the pervasiveness and flexibility of \acrshort{mlp}s, we trained \acrshort{mlp}s with different numbers of hidden layers with sigmoid activation function but without applying \acrlong{bn} and \acrlong{fe} at first. The evaluation results based on \acrshort{rmse} are given in \autoref{fig:rmse_vs_hidden}. Note that only the \acrshort{rmse} on the test set from new datasets is the proper indicator of the generalization performance (as described in section \ref{sec:dataset}). Looking at the test set performance shown in the orange curve in \autoref{fig:rmse_vs_hidden}, it is apparent that increasing the depth of the model further after two hidden layers surprisingly brings down the model performance. Similar observations were also reported in \citep{heConvolutionalNeuralNetworks2014,zeilerRectifiedLinearUnits2013,simonyanVeryDeepConvolutional2015}, where aggressively increasing the depth leads to saturated or degraded accuracy. Since two-hidden-layer \acrshort{mlp} already provided the best generalization performance, we chose the two-hidden-layer \acrshort{mlp} as the baseline.

It's also worth mentioning that we also kept computational resources in mind when choosing the baseline. Although a high-end GPU (NVIDIA 3090) currently did all the training and analyses, we plan to deploy the model on an edge computing device or on a real-time platform in the future. In fact, the two-hidden-layer \acrshort{mlp} is not necessarily always the best-performing model without computational limitations. One argument is that the shape of the test performance curve in \autoref{fig:rmse_vs_hidden} is a good match for the double-descent risk curve in \cite{belkinReconcilingModernMachinelearning2019}, which indicates that introducing more hidden layers could lead to even lower \acrshort{rmse} on the test set. Another clue lies in the training error. Except for the fact that it almost overlapped with the curve of "RMSE on the test set from random split" as explained in section \ref{sec:dataset}, the training error slightly rises rather than falls once the model size exceeds three hidden layers. This phenomenon where the training error increases as the model gets bigger and deeper is the degradation problem reported in \cite{heDeepResidualLearning2015} and \cite{srivastavaHighwayNetworks2015}. The degradation problem could be solved by techniques such as residual learning reformulation \citep{heDeepResidualLearning2015}. However, considering the tradeoff between the computational burden and model performance, as well as the fact that even this simple baseline already outperforms the traditional physical model (section \ref{sec:comparison}), we still prefer the two-hidden-layer \acrshort{mlp} as the baseline. To make the comparisons among the models below and the baseline fair, we limited the parameter numbers of all the following models to a roughly equivalent range of about 10K. Based on this guiding principle, we chose the real- and complex-valued \acrshort{cnn}s as illustrated in \autoref{fig:cnn} and discussed in section \ref{sec:methods}.
%
% Tables_ml
\begin{table}[ht]
    \centering
    \begin{threeparttable}
        \setlength\tabcolsep{4pt} % column space, default value: 6pt
        \begin{tabular}{c c c c c}
        \hline
            Framework   &   train \acrshort{rmse} & test \acrshort{rmse}    &   \acrshort{re}(\%)\\
            \hline
            \textbf{\acrshort{mlp}(baseline)}    &   1.99    &   \textbf{5.56}    &   \textbf{0.31} \\
            Physical model\tnote{*}  &   -   &   13.61  &   0.75\ \\ 
             \acrshort{rf}  &   \textbf{0.89}    &   44.35   &   2.44  \\ 
             \acrshort{gbdt}    &   4.77    &   38.56   &   2.12 \\ 
             \hline
        \end{tabular}
        \begin{tablenotes}
            \footnotesize
            \item[*] We evaluated the traditional physical model only on the test set since it doesn't need a training process (so its training \acrshort{rmse} is left blank). The \acrshort{rmse} of 13.61 mm, which corresponds to a \acrshort{re} of 0.75\%, is the average performance of the physical model on the test set. 
        \end{tablenotes}
    \end{threeparttable}
    \caption{Performance comparison of baseline \acrfull{mlp} (deep learning), \acrfull{rf} and \acrfull{gbdt} (classical machine learning), as well as the traditional physical model. For each model, the performance is represented as \acrfull{rmse} on the training and test set. The \acrfull{re}, which is the ratio between test \acrshort{rmse} and the cylinder stroke, is also listed.}
    \label{tab:ml}
\end{table}
\textbf{Baseline vs Classical.} We first compare the baseline \acrshort{mlp} model with the traditional physical model \citep{Scheidt2017} and classical machine learning models described in section \ref{sec:ml}. 

The physical model generally performs well in a mild and stable environment. However, its accuracy drops when dealing with our complex working scenarios and a wide range of temperature changes. We tested the performance of the physical model on the test set and calculated an average \acrshort{rmse} of \SI{13.61}{\milli\meter} obtained from measurements ranging from room temperature up to \SI{84}{\celsius}. For a cylinder stroke of \SI{1815}{\milli\meter}, it corresponds to an \acrshort{re} of 0.75\%, which is outperformed by the baseline model with a test \acrshort{rmse} as \SI{5.56}{\milli\meter}.

Both \acrshort{rf} and \acrshort{gbdt} suffer from poor generalization behavior in our case, which can be inferred from the massive gap between the train and test \acrshort{rmse}, especially for \acrshort{rf}. The performance of \acrshort{gbdt} is inferior to our baseline \acrshort{mlp} or even the traditional physical model. Considering that the models for both frameworks are already the best choice after extensive hyperparameter fine-tuning with 500 trials each, we thus confirm our hypothesis that it's the lack of well-designed features with domain knowledge that makes it difficult for classical machine learning methods to obtain comparable performance on our piston position task using scattering parameters. Thereafter we turned our attention to deep learning and only adapted \acrshort{ann} models to enhance their performance.

\textbf{\acrfull{bn}.} The first technique we applied to the baseline model was \acrshort{bn} \citep{ioffeBatchNormalizationAccelerating2015}. As can be seen from the first two lines of \autoref{tab:mc}, it helps improve the test \acrshort{rmse} by 30\% from 5.56 to 3.89 mm. However, this is not the case for \acrshort{cnn}, \acrshort{cvnn}, or models with \acrlong{fe} which hardly benefit from \acrshort{bn} in our case. 

\begin{figure}
    \centering
    \includegraphics[width=0.8\columnwidth]{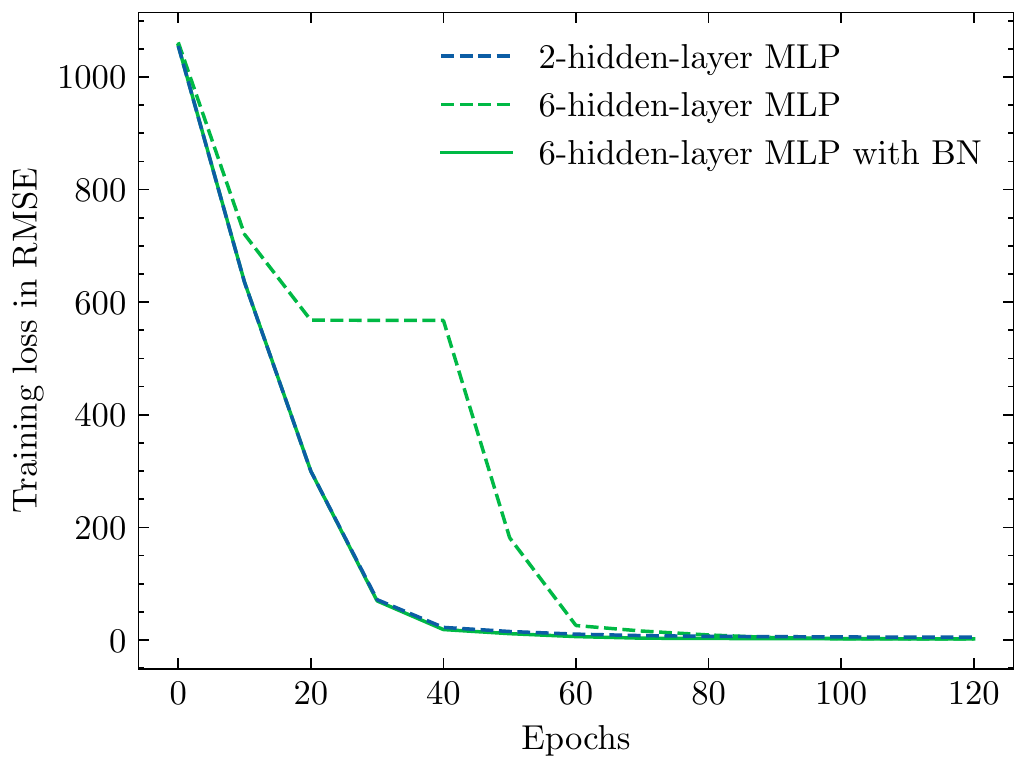}
    \caption{Training loss versus epochs for \acrfull{mlp} models with different numbers of hidden layers. The deep (6-hidden-layer) model suffers from an unstable training process due to the vanishing gradient issue. \acrfull{bn} help eliminate this problem, bringing the training of the deep model back to the normal situation, shown as the training process of a shallow model (2-hidden-layer \acrshort{mlp}).}
    \label{fig:loss}
\end{figure}

Another point worth mentioning is that \acrshort{bn} does help stabilize the training process for deep \acrshort{mlp}s. As can be seen in \autoref{fig:loss}, the 6-hidden-layer \acrshort{mlp} without \acrshort{bn} cease to advance for more than 20 epochs by showing a long plateau of training errors likely to be caused by vanishing gradients, especially for deeper models, while \acrshort{bn} eliminates this phenomenon, since it ensures forward propagated signals to have non-zero variances and therefore healthy norms of gradients during backpropagation.

\textbf{Activation Functions.} The choice of activation functions in deep networks has a significant effect on the training dynamics and task performance \citep{ramachandranSearchingActivationFunctions2017}. \autoref{tab:act} compares the generalization performance (\acrshort{me}, \acrshort{mae}, \acrshort{rmse}, and \acrshort{re} on the test set) of 2-hidden-layer batch-normalized \acrshort{mlp}s with four different activation functions. Previous research such as \cite{glorotDeepSparseRectifier2011} shows that ReLU generally performs better in most deep learning models, especially for image and text data. In our case, it converges faster too, but can not provide promising results in our case. Its variant, Leaky ReLU, exhibits similar behavior. Sigmoid, on the other hand, outperforms other activations. We believe this intriguing result is attributed to the fact that the sigmoid function, $f(z) = 1/(1+\exp(-z))$, is more in line with the physical characteristic of the input transmission $\bm{t}$ expressed in \mautoref{eq:euler,eq:t_of_l} since exponential function dominates. Components that are better reflections of the real physical world allow models to learn the relationship $L=f^{-1}(\bm{t})$ better, especially for shallow \acrshort{ann} modes like our baseline, where the representation learning ability is restricted by the depth and parameters of the model. This assumption also explains why SELU, a half-linear-half-exponential function, comes in between the more exponential sigmoid and the more linear ReLU. Because of the excellent performance of the sigmoid activation, we adopted it for all the following models.
%
% Tables_activation
\begin{table*}[h]
    \centering
    \begin{tabular}{c | c c |c c c c}
    \hline
        Framework & \acrshort{bn} & Activation & \acrshort{me} & \acrshort{mae} & \acrshort{rmse} & \acrshort{re}(\%) \\\hline
         \acrshort{mlp} & \checkmark & \textbf{Sigmoid} & -0.18 & \textbf{2.08} & \textbf{3.89} & \textbf{0.21} \\ 
         \acrshort{mlp} & \checkmark & SELU & -0.08 & 4.3 & 5.88 & 0.32 \\ 
         \acrshort{mlp} & \checkmark & ReLU & -0.06 & 4.00 & 5.96 & 0.33 \\ 
        \acrshort{mlp} & \checkmark & LeakyReLU & \textbf{0.00} & 4.79 & 8.3 & 0.46 \\ \hline
    \end{tabular}
    \caption{Performance comparison for models with different activation functions. All models in the table are 2-hidden-layer real-valued \acrshort{mlp}s with \acrfull{bn}. The only difference is their activation functions.}
    \label{tab:act}
\end{table*}
%
%
% Tables_model_comparison
\begin{table*}[h]
    \centering
    \begin{tabular}{c| c c c | c c c c}
    \hline
        Framework & \acrshort{bn} & \acrshort{fe} & \acrshort{cvnn} & \acrshort{me} & \acrshort{mae} & \acrshort{rmse} & \acrshort{re}(\%) \\\hline
        \acrshort{mlp} (baseline) & ~ & ~ & ~ & -0.1 & 2.04 & 5.56 & 0.31 \\
        \acrshort{mlp} & \checkmark & ~ & ~ & -0.18 & 2.08 & 3.89 & 0.21 \\
        \acrshort{mlp} & ~ & \checkmark & ~ & -0.11 & 1.24 & 2.5 & 0.14 \\
        \acrshort{mlp} & ~ & ~ & \checkmark & -0.07 & 0.95 & 1.53 & 0.08 \\
        \acrshort{mlp} & ~ & \textbf{\checkmark} & \textbf{\checkmark} & \textbf{-0.07} & \textbf{0.75} & \textbf{1.23} & \textbf{0.07} \\\hline
        \acrshort{cnn} & ~ & ~ & ~ & 0.34 & 1.27 & 2.32 & 0.10 \\
        \acrshort{cnn} & \checkmark & ~ & ~ & 0.82 & 1.69 & 2.35 & 0.13 \\
        \acrshort{cnn} & ~ & \checkmark & ~ & \textbf{0.11} & 0.8 & 1.47 & 0.08 \\
        \acrshort{cnn} & ~ & ~ & \checkmark & 0.45 & 1.02 & 1.53 & 0.08 \\
        \acrshort{cnn} & ~ & \textbf{\checkmark} & \textbf{\checkmark} & 0.18 & \textbf{0.74} & \textbf{1.01} & \textbf{0.06} \\\hline
        
    \end{tabular}
    \caption{Performance comparison for models with different architectures. Both frameworks, \acrshort{mlp} (\autoref{fig:mlp}) and \acrshort{cnn} (\autoref{fig:cnn}) are tested with techniques \acrfull{bn} and \acrfull{fe}. Check marks in the \acrfull{cvnn} column mean that we adopted complex-valued \acrfull{mlp} (\autoref{fig:cvmlp}) or complex-valued \acrfull{cnn} (\autoref{fig:cvcnn}) for the model in that specific row. The model performance is shown by \acrfull{me}, \acrfull{mae}, \acrfull{rmse} and \acrfull{re} based on \acrshort{rmse}.}
    \label{tab:mc}
\end{table*}

\textbf{Ablation Study.} Evaluation results of \acrshort{mlp}s and \acrshort{cnn}s are provided in \autoref{tab:mc}. The top and bottom parts of the table show models in these two frameworks separately. For models in both frameworks, \acrfull{bn}, \acrfull{fe}, and \acrfull{cvnn} were applied to them selectively. We kept only the promising models and omitted poorly performing ones. Since \acrshort{bn} doesn't perform well along with the other two techniques in our test, we always applied \acrshort{fe} and \acrshort{cvnn} without \acrshort{bn}. In the following paragraphs, we first focus on the \acrshort{mlp} part of the table to discuss the outcome of \acrshort{fe} and the impact of \acrshort{cvnn}. Finally, we compare \acrshort{mlp} and \acrshort{cnn} and relate the model performance to the physical characteristic of our data.

\textbf{\acrfull{fe}.} This simple technique is able to reduce the test \acrshort{rmse} by more than half from 5.56 to 2.5 mm, as shown by the first and the third row of \autoref{tab:mc}. As stated in section \ref{sec:fe}, we apply \acrshort{fe} with learnable weights in a multiplicative way and add the weighted codes to the original input. This also means the significant improvement by \acrshort{fe} was achieved by just adding $121 \times 2 = 242$ parameters to our baseline, which is a model with about 10K parameters in total. In other words, \acrshort{fe} boosts the model performance by 55\% at the cost of merely 2\% of more parameters. This result strongly supports our hypothesis that providing the model with frequency information will significantly improve the model's performance. It also suggests that we should build learning-based models as physically close to the real-world application as possible. As a side note, we also tested a variant of the standard positional encoding, which encodes the frequency by sinusoidal functions and directly adds to the input without any weighting process. This mode only slightly improves the model performance compared to the baseline. The reason lies in the fact that we are not able to determine in advance the specific effect of frequency on the input. Modeling this impact by a blindly chosen function, which only makes the frequencies unique instead of informative, turns out not to be a good design. Thus, we prefer the implementation with learnable weights.

\textbf{\acrfull{cvnn}.}
Besides real-valued \acrshort{mlp}s (with architecture illustrated in \autoref{fig:rvmlp} and performance given in the first three rows of section \ref{tab:mc}), \autoref{tab:mc} also displays the test performance of complex-valued \acrshort{mlp}s (with structure depicted in \autoref{fig:cvmlp}) indicated by check marks in the column \acrshort{cvnn}. As discussed in section \ref{sec:dl}, one layer of our \acrshort{cvnn} model consists of two standard layers representing the real and imaginary parts of complex-valued weights, respectively. We combine them in a way to mimic the calculation of complex numbers. To ensure fairness in comparing complex-valued and real-valued models, we carefully adjusted the depth and width for each \acrshort{cvnn} so that their total number of parameters was close to their real-valued counterparts (see section \ref{sec:complexity}). We conjecture that the \acrshort{cvnn} helps with the model performance due to the complex-valued characteristic of the model input and traditional physical model we would like to approximate. We confirm our hypothesis by comparing the first and fourth rows of the \autoref{tab:mc}. The complex-valued 2-hidden-layer \acrshort{mlp} significantly cuts down the test \acrshort{rmse} by over 72\% to 1.53 mm. This improvement comes only at a price of about 7\% of more model parameters. 

Incorporating \acrshort{fe} into \acrshort{cvnn} can further enhance model performance. As a result, the integration of these two components leads to the best performance for models based on the \acrshort{mlp} framework. The test RMSE drops by 78\% to 1.23 mm while the model is only 9\% bigger than the baseline. Compared to the fact that we have known from \autoref{fig:rmse_vs_hidden} that stacking layers blindly is, instead of an upgrade, but a downgrade of the model performance, the achievement by \acrshort{fe} and \acrshort{cvnn} is undoubtedly phenomenal.

\begin{figure}
    \centering
    \includegraphics[width=\columnwidth]{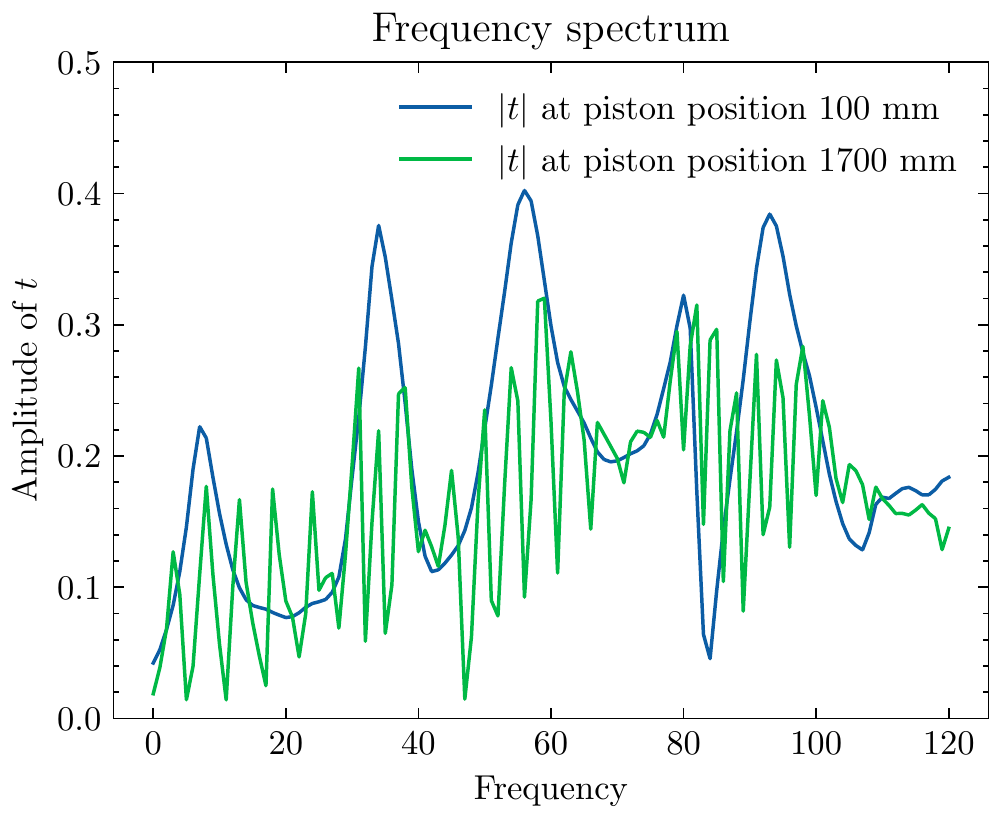}
    \caption{Frequency spectrum of transmission $\bm{t}$ at two different piston positions. The amplitudes of model input $\bm{t}$ at the shown two positions show clearly different local patterns, which are assumed to be easily detected by a \acrshort{cnn}.}
    \label{fig:freq_spec}
\end{figure}

\textbf{\acrshort{mlp} vs \acrshort{cnn}.} Compared to \acrshort{mlp}, \acrshort{cnn} is a better framework for our task. First, the enhancement due to \acrlong{fe} and \acrshort{cvnn} on \acrshort{cnn} is in line with the progress they made on \acrshort{mlp}. What really stands out about \acrshort{cnn}s (as listed in the \autoref{tab:mc}) is that their overall performance is consistently better than that of \acrshort{mlp}s in all cases. We believe the superiority of \acrshort{cnn} has two main reasons. Firstly, it is for the sake of the parameter sharing mechanism of \acrshort{cnn}, which enables it to contain more layers with the same number of model parameters. Therefore, limiting the number of model parameters results in the fact that \acrshort{cnn}s are able to take advantage of the extra depth. A deeper architecture provides an exponentially increased expressive capability and is more likely to learn rich distributed representations \citep{rumelhartLearningRepresentationsBackpropagating1986} of data, which is beneficial to the regression task of the final layer. More importantly, local connections between the frequencies also play an essential role here. For example, the position of the piston will have a local effect on the frequency spectrum of the input transmission $\bm{t}$. As shown in \autoref{fig:freq_spec}, the shape and form of the peaks and valleys in the spectrum are different when the corresponding piston position changes. We know from the physical model that the damping of the peak is frequency-dependent, and the location of the peak hinges on the piston position. Those local changes of peak patterns can be easily detected by a \acrshort{cnn} since its feature maps for each layer only focus on local patches. 

\textbf{The best model.} Finally, we propose our best-performing model: the complex-valued \acrshort{cnn} with \acrlong{fe}. It manages to reduce the test \acrshort{rmse} by 82\% compared to the baseline \acrshort{mlp} model. This achievement is entirely attributable to model design based on the physical characteristics of the system. We hope this will remind researchers working on projects related to scattering parameters to focus on the design of the deep learning model rather than simply adopting the standard model. Physical models remain at the core of the field, but the direct computation of physical quantities through equations that are not suitable for all scenarios may no longer be the optimal solution. Compared to the \acrshort{rmse} given by the traditional physical model, our dedicated best-performing deep learning model exhibits a reduction by a factor of 1/12, demonstrating a substantial improvement in accuracy and robustness considering the diversity of the datasets.

\subsection{Model complexity}
\label{sec:complexity}
While the previous sections focus on the model performance in terms of accuracy, in this section we shift our attention to speed and complexity. All test results regarding model complexity are given in \autoref{tab:complexity}.
%
% Tables_model_complexity
\begin{table*}[h]
    \centering
    \begin{threeparttable}
        \begin{tabular}{c|c c|c c c c}
            \hline
            Framework & \acrshort{fe} & \acrshort{cvnn}\tnote{*} & \makecell{Params\\(K)} & \makecell{Training\\ (s/epoch)} & \makecell{Inference\textsubscript{3090}\\(\si{\micro\second})\tnote{**}} & \makecell{Inference\textsubscript{NX}\\(\si{\micro\second})\tnote{**}}
            \\\hline
            \acrshort{mlp} &            &            & 8  & 8  & \multirow{6}{*}{$\approx$\num{2e-5}}     & 6 \\ 
            \acrshort{mlp} & \checkmark &            & 9  & 9  &                                        & 7 \\
            \acrshort{mlp} & \checkmark & \checkmark & 9  & 18 &                                        & 27 \\ 
            \acrshort{cnn} &            &            & 9  & 32 &                                        & 77 \\ 
            \acrshort{cnn} & \checkmark &            & 9  & 33 &                                        & 95 \\ 
            \acrshort{cnn} & \checkmark & \checkmark & 10 & 65 &                                        & 161 \\ \hline
        \end{tabular}
        \begin{tablenotes}
            \footnotesize
            \item[*] Instead of adopting torch.complex64 directly for the implementation of \acrshort{cvnn}s in this table, we used two torch.float32 tensors to replace the real and imaginary parts of a torch.complex64 weight, respectively. We found that torch.complex64 significantly reduces the training and inference speed.
            \item[**]  \SI{1}{\micro\second} = \SI{1e-6}{\second}
        \end{tablenotes}
    \end{threeparttable}
    
    \caption{Model complexity for proposed deep learning models. The number of model parameters, training, and inference speed of \acrshort{mlp} and \acrshort{cnn} models with their \acrfull{fe} or \acrfull{cvnn} variants are listed. The number of parameters for all models is kept around 10K by design. The training speed is recorded on an NVIDIA GeForce RTX 3090 and in the form of seconds per epoch with each epoch containing 411K data points. The inference speed is tested on the same RTX 3090 (the second to last column) and in addition on the CPU of Jetson Xavier NX in the lowest power mode (the last column).}
    \label{tab:complexity}
\end{table*}
As previously mentioned, we intentionally limit the parameters of the model to 10K to preserve the potential for deployment on an edge computing device. The number of parameters for our deep learning models is specified in the fourth column of \autoref{tab:complexity}. Note that our most accurate model (in the last row) has slightly more parameters than the other models.  However, this is not the reason for its superior accuracy.  As shown in \autoref{fig:rmse_vs_hidden}, blindly stacking model parameters can deteriorate the model performance. This is also the reason that the baseline \acrshort{mlp} only has 8K parameters, as the 10K version was found to be less accurate. Therefore, the enhancement should be attributed to better model design considering the physical characteristics of the system, i.e. \acrshort{cnn}, \acrshort{fe}, and \acrlong{cvnn}, rather than the extra parameters. 

The training speed is recorded on an NVIDIA GeForce RTX 3090 and in the form of seconds per epoch with each epoch containing 411K data points. We note that the training time for \acrshort{cnn} is longer than \acrshort{mlp}. This is because that \acrshort{cnn} with the same parameters will perform more operations than \acrshort{mlp} because of \acrshort{cnn}'s parameter-sharing mechanism.
Similarly, \acrshort{cvnn} is slower than real-valued \acrshort{ann} due to extra computation and IO cost for the complex domain (\autoref{fig:cvnn}). Also, note that the low-level implementation of data type could also lead to different training speeds. We implemented \acrshort{cvnn} both in "torch.complex64" and "torch.float32". The former can directly implement a complex-valued tensor, and the latter replaces the real and imaginary parts of a complex-valued tensor with two real-valued tensors. The two implementations are mathematically equivalent, but we find that torch.complex64 significantly reduces the training and inference speed (e.g. for about 50\%, from 65 to 94 s/epoch for the training time on RTX 3090 and from 161 to 243 \si{\micro\second} for the inference time on Jetson Xavier NX, in the case of the \acrshort{cnn} framework in the last row). 
 
We also tested the inference speed of our models. Our test results on the 3090 were unstable and convergent, as the speed of the calculations was too fast, leading to too pronounced random fluctuations. Our test results on all models are around \num{2e-5} $\mu$s. To represent the potential of the model to operate on edge computing devices, we tested the inference speed on the CPU of a Jeston Xavier NX at the lowest power mode (10 W, 2 cores). Here, the inference time demonstrates a similar pattern to the training time since they are both strongly influenced by the actual operations performed in the hardware. Note that the unit of the inference speed is microsecond \si{\micro\second} where\SI{1}{\micro\second} = \SI{1e-6}{\second}. This implies that even the slowest model is capable of performing more than 6000 inferences per second, sufficient for real-time applications.

\section{Conclusion}
\label{sec:conclusion}

In this paper, deep learning models were proposed and evaluated for the piston position detection task inside a cylinder. The input to our models, which is measured by a device named LiView, is a mathematical expression of scattering parameters of a three-port network. The training and test set are carefully chosen to accurately reflect the generalization performance of our models. Considering the possibility of future integration of our model into the LiView electronic unit, the number of parameters was restricted to about 10K for all models. It is demonstrated that deep learning models (\acrshort{mlp} and \acrshort{cnn}) consistently outperformed classical machine learning methods (\acrshort{rf} and \acrshort{gbdt}) and the traditional physical approach. Further improvements were made by incorporating \acrlong{fe} and \acrshort{cvnn} to the deep learning models, resulting in 55\% and 72\% improvement in generalization performance at the cost of merely 2\% and 7\% extra model parameters, respectively. The best-performing model consisting of the three winning components, a complex-valued \acrshort{cnn} with \acrshort{fe}, managed to plummet the error to hardly 1/12 compared to the traditional physical model. 

The potentials of combining deep learning and scattering parameters are far from being limited to the position detection task. We plan to extend the paradigm to predictive maintenance and investigate oil permittivity problems. We are excited about the future of a deep learning-based LiView.

% % list of acronyms
% \glsaddall
% \printglossary[type=\acronymtype,title=Acronyms]

%% If you have bibdatabase file and want bibtex to generate the
%% bibitems, please use
%%
\bibliographystyle{elsarticle-harv} 
\bibliography{ref}

\begin{thebibliography}{65}
\expandafter\ifx\csname natexlab\endcsname\relax\def\natexlab#1{#1}\fi
\providecommand{\url}[1]{\texttt{#1}}
\providecommand{\href}[2]{#2}
\providecommand{\path}[1]{#1}
\providecommand{\DOIprefix}{doi:}
\providecommand{\ArXivprefix}{arXiv:}
\providecommand{\URLprefix}{URL: }
\providecommand{\Pubmedprefix}{pmid:}
\providecommand{\doi}[1]{\href{http://dx.doi.org/#1}{\path{#1}}}
\providecommand{\Pubmed}[1]{\href{pmid:#1}{\path{#1}}}
\providecommand{\bibinfo}[2]{#2}
\ifx\xfnm\relax \def\xfnm[#1]{\unskip,\space#1}\fi
%Type = Article
\bibitem[{Akhavanhejazi et~al.(2011)Akhavanhejazi, Gharehpetian, {Faraji-dana},
  Moradi, Mohammadi and Alehoseini}]{akhavanhejaziNewOnlineMonitoring2011}
\bibinfo{author}{Akhavanhejazi, M.}, \bibinfo{author}{Gharehpetian, G.},
  \bibinfo{author}{{Faraji-dana}, R.}, \bibinfo{author}{Moradi, G.},
  \bibinfo{author}{Mohammadi, M.}, \bibinfo{author}{Alehoseini, H.},
  \bibinfo{year}{2011}.
\newblock \bibinfo{title}{A new on-line monitoring method of transformer
  winding axial displacement based on measurement of scattering parameters and
  decision tree}.
\newblock \bibinfo{journal}{Expert Systems with Applications}
  \bibinfo{volume}{38}, \bibinfo{pages}{8886--8893}.
\newblock \DOIprefix\doi{10.1016/j.eswa.2011.01.100}.
%Type = Misc
\bibitem[{Akiba et~al.(2019)Akiba, Sano, Yanase, Ohta and
  Koyama}]{akibaOptunaNextgenerationHyperparameter2019}
\bibinfo{author}{Akiba, T.}, \bibinfo{author}{Sano, S.},
  \bibinfo{author}{Yanase, T.}, \bibinfo{author}{Ohta, T.},
  \bibinfo{author}{Koyama, M.}, \bibinfo{year}{2019}.
\newblock \bibinfo{title}{Optuna: {{A Next-generation Hyperparameter
  Optimization Framework}}}.
\newblock \DOIprefix\doi{10.48550/arXiv.1907.10902},
  \href{http://arxiv.org/abs/1907.10902}{{\tt arXiv:1907.10902}}.
%Type = Article
\bibitem[{Bader et~al.(2021)Bader, Haddad, Kallel, Hassine, Amara and
  Kanoun}]{baderIdentificationCommunicationCables2021}
\bibinfo{author}{Bader, O.}, \bibinfo{author}{Haddad, D.},
  \bibinfo{author}{Kallel, A.Y.}, \bibinfo{author}{Hassine, T.},
  \bibinfo{author}{Amara, N.E.B.}, \bibinfo{author}{Kanoun, O.},
  \bibinfo{year}{2021}.
\newblock \bibinfo{title}{Identification of {{Communication Cables Based}} on
  {{Scattering Parameters}} and a {{Support Vector Machine Algorithm}}}.
\newblock \bibinfo{journal}{IEEE Sensors Letters} \bibinfo{volume}{5},
  \bibinfo{pages}{1--4}.
\newblock \DOIprefix\doi{10.1109/LSENS.2021.3087539}.
%Type = Article
\bibitem[{Barbosa et~al.(2019)Barbosa, Tarrago, {de Medeiros}, {de Melo}, {da
  Silva Lourenco Novo}, Coutinho, Alves, Lott~Neto, Gama and {dos
  Santos}}]{barbosaMachineLearningApproach2019}
\bibinfo{author}{Barbosa, D.C.P.}, \bibinfo{author}{Tarrago, V.L.},
  \bibinfo{author}{{de Medeiros}, L.H.A.}, \bibinfo{author}{{de Melo}, M.T.},
  \bibinfo{author}{{da Silva Lourenco Novo}, L.R.G.},
  \bibinfo{author}{Coutinho, M.d.S.}, \bibinfo{author}{Alves, M.M.},
  \bibinfo{author}{Lott~Neto, H.B.D.T.}, \bibinfo{author}{Gama, P.H.R.P.},
  \bibinfo{author}{{dos Santos}, R.G.M.}, \bibinfo{year}{2019}.
\newblock \bibinfo{title}{Machine {{Learning Approach}} to {{Detect Faults}} in
  {{Anchor Rods}} of {{Power Transmission Lines}}}.
\newblock \bibinfo{journal}{IEEE Antennas and Wireless Propagation Letters}
  \bibinfo{volume}{18}, \bibinfo{pages}{2335--2339}.
\newblock \DOIprefix\doi{10.1109/LAWP.2019.2932052}.
%Type = Article
\bibitem[{Bassey et~al.(2021)Bassey, Qian and
  Li}]{basseySurveyComplexValuedNeural2021}
\bibinfo{author}{Bassey, J.}, \bibinfo{author}{Qian, L.}, \bibinfo{author}{Li,
  X.}, \bibinfo{year}{2021}.
\newblock \bibinfo{title}{A {{Survey}} of {{Complex-Valued Neural Networks}}}
  \DOIprefix\doi{10.48550/arxiv.2101.12249},
  \href{http://arxiv.org/abs/2101.12249}{{\tt arXiv:2101.12249}}.
%Type = Article
\bibitem[{Belkin et~al.(2019)Belkin, Hsu, Ma and
  Mandal}]{belkinReconcilingModernMachinelearning2019}
\bibinfo{author}{Belkin, M.}, \bibinfo{author}{Hsu, D.}, \bibinfo{author}{Ma,
  S.}, \bibinfo{author}{Mandal, S.}, \bibinfo{year}{2019}.
\newblock \bibinfo{title}{Reconciling modern machine-learning practice and the
  classical bias\textendash variance trade-off}.
\newblock \bibinfo{journal}{Proceedings of the National Academy of Sciences}
  \bibinfo{volume}{116}, \bibinfo{pages}{15849--15854}.
\newblock \DOIprefix\doi{10/gf5dmw}.
%Type = Article
\bibitem[{Bertolini et~al.(2021)Bertolini, Mezzogori, Neroni and
  Zammori}]{bertoliniMachineLearningIndustrial2021}
\bibinfo{author}{Bertolini, M.}, \bibinfo{author}{Mezzogori, D.},
  \bibinfo{author}{Neroni, M.}, \bibinfo{author}{Zammori, F.},
  \bibinfo{year}{2021}.
\newblock \bibinfo{title}{Machine {{Learning}} for industrial applications:
  {{A}} comprehensive literature review}.
\newblock \bibinfo{journal}{Expert Systems with Applications}
  \bibinfo{volume}{175}, \bibinfo{pages}{114820}.
\newblock \DOIprefix\doi{10.1016/j.eswa.2021.114820}.
%Type = Book
\bibitem[{Bishop and Nasrabadi(2006)}]{bishop2006pattern}
\bibinfo{author}{Bishop, C.M.}, \bibinfo{author}{Nasrabadi, N.M.},
  \bibinfo{year}{2006}.
\newblock \bibinfo{title}{Pattern recognition and machine learning}.
  volume~\bibinfo{volume}{4}.
\newblock \bibinfo{publisher}{Springer}.
%Type = Misc
\bibitem[{Braun et~al.(2014-01-17)Braun, Engler, Cremer and Nannen}]{Braun2014}
\bibinfo{author}{Braun, S.}, \bibinfo{author}{Engler, A.},
  \bibinfo{author}{Cremer, R.}, \bibinfo{author}{Nannen, I.},
  \bibinfo{year}{2014-01-17}.
\newblock \bibinfo{title}{Verfahren zur bestimmung der kolbenposition einer
  kolbenzylindereinheit und kolbenzylindereinheit}.
\newblock \URLprefix \url{https://patents.google.com/patent/EP2759715A1}.
%Type = Misc
\bibitem[{Braun et~al.(2017-04-05)Braun, Scheidt, Kipp and
  Leutenegger}]{Scheidt2017}
\bibinfo{author}{Braun, S.}, \bibinfo{author}{Scheidt, M.},
  \bibinfo{author}{Kipp, M.}, \bibinfo{author}{Leutenegger, P.},
  \bibinfo{year}{2017-04-05}.
\newblock \bibinfo{title}{Vorrichtung und verfahren zur positionsbestimmung
  eines zylinderkolbens}.
\newblock \URLprefix
  \url{https://patentimages.storage.googleapis.com/cf/71/92/db2668332545b7/DE102015012799A1.pdf}.
%Type = Article
\bibitem[{Breiman(2001)}]{breimanRandomForests2001}
\bibinfo{author}{Breiman, L.}, \bibinfo{year}{2001}.
\newblock \bibinfo{title}{Random {{Forests}}}.
\newblock \bibinfo{journal}{Machine Learning} \bibinfo{volume}{45},
  \bibinfo{pages}{5--32}.
\newblock \DOIprefix\doi{10/d8zjwq}.
%Type = Book
\bibitem[{Breiman et~al.(1984)Breiman, Friedman, Olshen and
  Stone}]{reason:BreFriOlsSto84a}
\bibinfo{author}{Breiman, L.}, \bibinfo{author}{Friedman, J.H.},
  \bibinfo{author}{Olshen, R.A.}, \bibinfo{author}{Stone, C.J.},
  \bibinfo{year}{1984}.
\newblock \bibinfo{title}{Classification and Regression Trees}.
\newblock \bibinfo{publisher}{Wadsworth and Brooks},
  \bibinfo{address}{Monterey, CA}.
%Type = Misc
\bibitem[{Büchler et~al.(2011-06-30)Büchler, Fericean, Dorneich and
  Fritton}]{Balluff2}
\bibinfo{author}{Büchler, J.}, \bibinfo{author}{Fericean, S.},
  \bibinfo{author}{Dorneich, A.}, \bibinfo{author}{Fritton, M.},
  \bibinfo{year}{2011-06-30}.
\newblock \bibinfo{title}{Verfahren und vorrichtung zur ermittlung der position
  eines kolbens eines kolbenzylinders mit mikrowellen}.
\newblock \URLprefix \url{https://patents.google.com/patent/DE102009055445A1}.
%Type = Inproceedings
\bibitem[{Chen and Guestrin(2016a)}]{chenXGBoostScalableTree2016}
\bibinfo{author}{Chen, T.}, \bibinfo{author}{Guestrin, C.},
  \bibinfo{year}{2016}a.
\newblock \bibinfo{title}{{{XGBoost}}: {{A Scalable Tree Boosting System}}},
  in: \bibinfo{booktitle}{Proceedings of the 22nd {{ACM SIGKDD International
  Conference}} on {{Knowledge Discovery}} and {{Data Mining}}}, pp.
  \bibinfo{pages}{785--794}.
\newblock \DOIprefix\doi{10/gdp84q},
  \href{http://arxiv.org/abs/1603.02754}{{\tt arXiv:1603.02754}}.
%Type = Article
\bibitem[{Chen and Guestrin(2016b)}]{DBLP:journals/corr/ChenG16}
\bibinfo{author}{Chen, T.}, \bibinfo{author}{Guestrin, C.},
  \bibinfo{year}{2016}b.
\newblock \bibinfo{title}{Xgboost: {A} scalable tree boosting system}.
\newblock \bibinfo{journal}{CoRR} \bibinfo{volume}{abs/1603.02754}.
\newblock \URLprefix \url{http://arxiv.org/abs/1603.02754},
  \href{http://arxiv.org/abs/1603.02754}{{\tt arXiv:1603.02754}}.
%Type = Misc
\bibitem[{Dorogush et~al.(2018)Dorogush, Ershov and
  Gulin}]{dorogushCatBoostGradientBoosting2018}
\bibinfo{author}{Dorogush, A.V.}, \bibinfo{author}{Ershov, V.},
  \bibinfo{author}{Gulin, A.}, \bibinfo{year}{2018}.
\newblock \bibinfo{title}{{{CatBoost}}: Gradient boosting with categorical
  features support}.
\newblock \href{http://arxiv.org/abs/1810.11363}{{\tt arXiv:1810.11363}}.
%Type = Misc
\bibitem[{Fend et~al.(2011-06-30)Fend, Audergon, Fritton and
  Dorneich}]{Balluff1}
\bibinfo{author}{Fend, N.}, \bibinfo{author}{Audergon, L.},
  \bibinfo{author}{Fritton, M.}, \bibinfo{author}{Dorneich, A.},
  \bibinfo{year}{2011-06-30}.
\newblock \bibinfo{title}{Verfahren zur bestimmung der position eines kolbens
  eines kolbenzylinders und mikrowellen-sensorvorrichtung}.
\newblock \URLprefix \url{https://patents.google.com/patent/DE102009055363A1}.
%Type = Article
\bibitem[{Frazier et~al.(2022)Frazier, Jr, Anlage and
  Ott}]{frazierDeepLearningEstimationComplex2022}
\bibinfo{author}{Frazier, B.W.}, \bibinfo{author}{Jr, T.M.A.},
  \bibinfo{author}{Anlage, S.M.}, \bibinfo{author}{Ott, E.},
  \bibinfo{year}{2022}.
\newblock \bibinfo{title}{Deep-{{Learning Estimation}} of {{Complex Reverberant
  Wave Fields}} with a {{Programmable Metasurface}}} ,
  \bibinfo{pages}{30}\DOIprefix\doi{10.1103/PhysRevApplied.17.024027}.
%Type = Inproceedings
\bibitem[{Glorot and Bengio(2010)}]{glorotUnderstandingDifficultyTraining2010}
\bibinfo{author}{Glorot, X.}, \bibinfo{author}{Bengio, Y.},
  \bibinfo{year}{2010}.
\newblock \bibinfo{title}{Understanding the difficulty of training deep
  feedforward neural networks}, in: \bibinfo{booktitle}{Proceedings of the
  {{Thirteenth International Conference}} on {{Artificial Intelligence}} and
  {{Statistics}}}, \bibinfo{publisher}{{JMLR Workshop and Conference
  Proceedings}}. pp. \bibinfo{pages}{249--256}.
%Type = Inproceedings
\bibitem[{Glorot et~al.(2011)Glorot, Bordes and
  Bengio}]{glorotDeepSparseRectifier2011}
\bibinfo{author}{Glorot, X.}, \bibinfo{author}{Bordes, A.},
  \bibinfo{author}{Bengio, Y.}, \bibinfo{year}{2011}.
\newblock \bibinfo{title}{Deep {{Sparse Rectifier Neural Networks}}}, in:
  \bibinfo{booktitle}{Proceedings of the {{Fourteenth International
  Conference}} on {{Artificial Intelligence}} and {{Statistics}}},
  \bibinfo{publisher}{{JMLR Workshop and Conference Proceedings}}. pp.
  \bibinfo{pages}{315--323}.
%Type = Book
\bibitem[{Goodfellow et~al.(2016)Goodfellow, Bengio and
  Courville}]{Goodfellow-et-al-2016}
\bibinfo{author}{Goodfellow, I.}, \bibinfo{author}{Bengio, Y.},
  \bibinfo{author}{Courville, A.}, \bibinfo{year}{2016}.
\newblock \bibinfo{title}{Deep Learning}.
\newblock \bibinfo{publisher}{MIT Press}.
\newblock \bibinfo{note}{\url{http://www.deeplearningbook.org}}.
%Type = Article
\bibitem[{Gupta(2020)}]{guptaHandMovementClassification}
\bibinfo{author}{Gupta, S.H.}, \bibinfo{year}{2020}.
\newblock \bibinfo{title}{Hand {{Movement Classification}} from {{Measured
  Scattering Parameters}} using {{Deep Convolutional Neural Network}}} ,
  \bibinfo{pages}{13}\DOIprefix\doi{10.1016/j.measurement.2019.107258}.
%Type = Misc
\bibitem[{He and Sun(2014)}]{heConvolutionalNeuralNetworks2014}
\bibinfo{author}{He, K.}, \bibinfo{author}{Sun, J.}, \bibinfo{year}{2014}.
\newblock \bibinfo{title}{Convolutional {{Neural Networks}} at {{Constrained
  Time Cost}}}.
\newblock \href{http://arxiv.org/abs/1412.1710}{{\tt arXiv:1412.1710}}.
%Type = Misc
\bibitem[{He et~al.(2015)He, Zhang, Ren and Sun}]{heDeepResidualLearning2015}
\bibinfo{author}{He, K.}, \bibinfo{author}{Zhang, X.}, \bibinfo{author}{Ren,
  S.}, \bibinfo{author}{Sun, J.}, \bibinfo{year}{2015}.
\newblock \bibinfo{title}{Deep {{Residual Learning}} for {{Image
  Recognition}}}.
\newblock \DOIprefix\doi{10.48550/arXiv.1512.03385},
  \href{http://arxiv.org/abs/1512.03385}{{\tt arXiv:1512.03385}}.
%Type = Article
\bibitem[{Hejazi et~al.(2011)Hejazi, Gharehpetian, Moradi, Alehosseini and
  Mohammadi}]{hejaziOnlineMonitoringTransformer2011}
\bibinfo{author}{Hejazi, M.A.}, \bibinfo{author}{Gharehpetian, G.B.},
  \bibinfo{author}{Moradi, G.}, \bibinfo{author}{Alehosseini, H.A.},
  \bibinfo{author}{Mohammadi, M.}, \bibinfo{year}{2011}.
\newblock \bibinfo{title}{Online monitoring of transformer winding axial
  displacement and its extent using scattering parameters and k-nearest
  neighbour method}.
\newblock \bibinfo{journal}{IET Gener. Transm. Distrib.} \bibinfo{volume}{5},
  \bibinfo{pages}{9}.
\newblock \DOIprefix\doi{10.1049/iet-gtd.2010.0802}.
%Type = Article
\bibitem[{Hien and Hong(2021)}]{hienMaterialThicknessClassification2021}
\bibinfo{author}{Hien, P.T.}, \bibinfo{author}{Hong, I.P.},
  \bibinfo{year}{2021}.
\newblock \bibinfo{title}{Material {{Thickness Classification Using Scattering
  Parameters}}, {{Dielectric Constants}}, and {{Machine Learning}}}.
\newblock \bibinfo{journal}{Applied Sciences} \bibinfo{volume}{11},
  \bibinfo{pages}{10682}.
\newblock \DOIprefix\doi{10.3390/app112210682}.
%Type = Book
\bibitem[{Hirose(2012)}]{hiroseComplexValuedNeuralNetworks2012}
\bibinfo{author}{Hirose, A.}, \bibinfo{year}{2012}.
\newblock \bibinfo{title}{Complex-{{Valued Neural Networks}}}. volume
  \bibinfo{volume}{400} of \textit{\bibinfo{series}{Studies in {{Computational
  Intelligence}}}}.
\newblock \bibinfo{publisher}{{Springer Berlin Heidelberg}},
  \bibinfo{address}{{Berlin, Heidelberg}}.
\newblock \DOIprefix\doi{10.1007/978-3-642-27632-3}.
%Type = Article
\bibitem[{Hornik et~al.(1989)Hornik, Stinchcombe and
  White}]{hornikMultilayerFeedforwardNetworks1989}
\bibinfo{author}{Hornik, K.}, \bibinfo{author}{Stinchcombe, M.},
  \bibinfo{author}{White, H.}, \bibinfo{year}{1989}.
\newblock \bibinfo{title}{Multilayer feedforward networks are universal
  approximators}.
\newblock \bibinfo{journal}{Neural Networks} \bibinfo{volume}{2},
  \bibinfo{pages}{359--366}.
\newblock \DOIprefix\doi{10/c4wch5}.
%Type = Article
\bibitem[{Husby et~al.(2019)Husby, Myrvoll and
  Knudsen}]{husbyEddyCurrentDuplex}
\bibinfo{author}{Husby, K.}, \bibinfo{author}{Myrvoll, T.A.},
  \bibinfo{author}{Knudsen, O.{\O}.}, \bibinfo{year}{2019}.
\newblock \bibinfo{title}{Eddy {{Current}} duplex coating thickness
  {{Non-Destructive Evaluation}} augmented by {{VNA}} scattering parameter
  theory and {{Machine Learning}}} ,
  \bibinfo{pages}{6}\DOIprefix\doi{10.1109/SAS.2019.8706003}.
%Type = Misc
\bibitem[{Ioffe and Szegedy(2015)}]{ioffeBatchNormalizationAccelerating2015}
\bibinfo{author}{Ioffe, S.}, \bibinfo{author}{Szegedy, C.},
  \bibinfo{year}{2015}.
\newblock \bibinfo{title}{Batch {{Normalization}}: {{Accelerating Deep Network
  Training}} by {{Reducing Internal Covariate Shift}}}.
\newblock \DOIprefix\doi{10.48550/arXiv.1502.03167},
  \href{http://arxiv.org/abs/1502.03167}{{\tt arXiv:1502.03167}}.
%Type = Article
\bibitem[{Kang et~al.(2020)Kang, Catal and
  Tekinerdogan}]{kangMachineLearningApplications2020}
\bibinfo{author}{Kang, Z.}, \bibinfo{author}{Catal, C.},
  \bibinfo{author}{Tekinerdogan, B.}, \bibinfo{year}{2020}.
\newblock \bibinfo{title}{Machine learning applications in production lines:
  {{A}} systematic literature review}.
\newblock \bibinfo{journal}{Computers \& Industrial Engineering}
  \bibinfo{volume}{149}, \bibinfo{pages}{106773}.
\newblock \DOIprefix\doi{10.1016/j.cie.2020.106773}.
%Type = Inproceedings
\bibitem[{Ke et~al.(2017)Ke, Meng, Finley, Wang, Chen, Ma, Ye and
  Liu}]{keLightGBMHighlyEfficient2017}
\bibinfo{author}{Ke, G.}, \bibinfo{author}{Meng, Q.}, \bibinfo{author}{Finley,
  T.}, \bibinfo{author}{Wang, T.}, \bibinfo{author}{Chen, W.},
  \bibinfo{author}{Ma, W.}, \bibinfo{author}{Ye, Q.}, \bibinfo{author}{Liu,
  T.Y.}, \bibinfo{year}{2017}.
\newblock \bibinfo{title}{{{LightGBM}}: {{A Highly Efficient Gradient Boosting
  Decision Tree}}}, in: \bibinfo{booktitle}{Advances in {{Neural Information
  Processing Systems}}}, \bibinfo{publisher}{{Curran Associates, Inc.}}
%Type = Article
\bibitem[{Kingma and Ba(2017)}]{diederikp.kingmaAdamMethodStochastic2017}
\bibinfo{author}{Kingma, D.P.}, \bibinfo{author}{Ba, J.}, \bibinfo{year}{2017}.
\newblock \bibinfo{title}{Adam: {{A Method}} for {{Stochastic Optimization}}}
  \DOIprefix\doi{10.48550/arxiv.1412.6980}.
%Type = Article
\bibitem[{Klambauer et~al.(2017)Klambauer, Unterthiner, Mayr and
  Hochreiter}]{gunterklambauerSelfNormalizingNeuralNetworks2017}
\bibinfo{author}{Klambauer, G.}, \bibinfo{author}{Unterthiner, T.},
  \bibinfo{author}{Mayr, A.}, \bibinfo{author}{Hochreiter, S.},
  \bibinfo{year}{2017}.
\newblock \bibinfo{title}{Self-{{Normalizing Neural Networks}}}
  \DOIprefix\doi{10.48550/arxiv.1706.02515}.
%Type = Article
\bibitem[{LeCun et~al.(2015)LeCun, Bengio and Hinton}]{lecunDeepLearning2015}
\bibinfo{author}{LeCun, Y.}, \bibinfo{author}{Bengio, Y.},
  \bibinfo{author}{Hinton, G.}, \bibinfo{year}{2015}.
\newblock \bibinfo{title}{Deep learning}.
\newblock \bibinfo{journal}{Nature} \bibinfo{volume}{521},
  \bibinfo{pages}{436--444}.
\newblock \DOIprefix\doi{10/bmqp}.
%Type = Article
\bibitem[{LeCun et~al.(1989)LeCun, Boser, Denker, Henderson, Howard, Hubbard
  and Jackel}]{lecunBackpropagationAppliedHandwritten1989}
\bibinfo{author}{LeCun, Y.}, \bibinfo{author}{Boser, B.},
  \bibinfo{author}{Denker, J.S.}, \bibinfo{author}{Henderson, D.},
  \bibinfo{author}{Howard, R.E.}, \bibinfo{author}{Hubbard, W.},
  \bibinfo{author}{Jackel, L.D.}, \bibinfo{year}{1989}.
\newblock \bibinfo{title}{Backpropagation {{Applied}} to {{Handwritten Zip Code
  Recognition}}}.
\newblock \bibinfo{journal}{Neural Computation} \bibinfo{volume}{1},
  \bibinfo{pages}{541--551}.
\newblock \DOIprefix\doi{10/bknd8g}.
%Type = Misc
\bibitem[{Leutenegger et~al.(2017-12-15)Leutenegger, Zinner, Scheidt and
  Stucke}]{Leutenegger2017}
\bibinfo{author}{Leutenegger, P.}, \bibinfo{author}{Zinner, T.},
  \bibinfo{author}{Scheidt, D.M.}, \bibinfo{author}{Stucke, M.},
  \bibinfo{year}{2017-12-15}.
\newblock \bibinfo{title}{Hochdruckdurchf{\"u}hrung zur durchf{\"u}hrung eines
  koaxialkabels in einen hochdruckbereich}.
\newblock \URLprefix \url{https://patents.google.com/patent/EP3349316A1}.
%Type = Article
\bibitem[{Loshchilov and Hutter(2017a)}]{loshchilovDecoupledWeightDecay2017}
\bibinfo{author}{Loshchilov, I.}, \bibinfo{author}{Hutter, F.},
  \bibinfo{year}{2017}a.
\newblock \bibinfo{title}{Decoupled weight decay regularization}.
\newblock \bibinfo{journal}{arXiv preprint arXiv:1711.05101}
  \DOIprefix\doi{10.48550/arxiv.1711.05101},
  \href{http://arxiv.org/abs/1711.05101}{{\tt arXiv:1711.05101}}.
%Type = Article
\bibitem[{Loshchilov and Hutter(2017b)}]{loshchilovSGDRStochasticGradient2017}
\bibinfo{author}{Loshchilov, I.}, \bibinfo{author}{Hutter, F.},
  \bibinfo{year}{2017}b.
\newblock \bibinfo{title}{{{SGDR}}: {{Stochastic Gradient Descent}} with {{Warm
  Restarts}}}.
\newblock \bibinfo{journal}{arXiv:1608.03983 [cs, math]}
  \DOIprefix\doi{10/gp345z}, \href{http://arxiv.org/abs/1608.03983}{{\tt
  arXiv:1608.03983}}.
%Type = Book
\bibitem[{Ludwig and Bretchko(2000)}]{ludwig2000}
\bibinfo{author}{Ludwig, R.}, \bibinfo{author}{Bretchko, P.},
  \bibinfo{year}{2000}.
\newblock \bibinfo{title}{RF Circuit Design: Theory and Applications}.
\newblock \bibinfo{publisher}{Prentice Hall}.
%Type = Article
\bibitem[{Maas et~al.(2013)Maas, Hannun and
  Ng}]{maasRectifierNonlinearitiesImprove}
\bibinfo{author}{Maas, A.L.}, \bibinfo{author}{Hannun, A.Y.},
  \bibinfo{author}{Ng, A.Y.}, \bibinfo{year}{2013}.
\newblock \bibinfo{title}{Rectifier {{Nonlinearities Improve Neural Network
  Acoustic Models}}} , \bibinfo{pages}{6}.
%Type = Article
\bibitem[{Matth\`es et~al.(2021)Matth\`es, Bromberg, de~Rosny and
  Popoff}]{PhysRevX.11.021060}
\bibinfo{author}{Matth\`es, M.W.}, \bibinfo{author}{Bromberg, Y.},
  \bibinfo{author}{de~Rosny, J.}, \bibinfo{author}{Popoff, S.M.},
  \bibinfo{year}{2021}.
\newblock \bibinfo{title}{Learning and avoiding disorder in multimode fibers}.
\newblock \bibinfo{journal}{Phys. Rev. X} \bibinfo{volume}{11},
  \bibinfo{pages}{021060}.
\newblock \URLprefix \url{https://link.aps.org/doi/10.1103/PhysRevX.11.021060},
  \DOIprefix\doi{10.1103/PhysRevX.11.021060}.
%Type = Misc
\bibitem[{Morgan(1994-06-28)}]{Caterpillar1994}
\bibinfo{author}{Morgan, D.E.}, \bibinfo{year}{1994-06-28}.
\newblock \bibinfo{title}{Linear position sensor with means to eliminate
  spurians harmonic detections}.
\newblock \URLprefix \url{https://patents.google.com/patent/US5325063A}.
%Type = Article
\bibitem[{Nair and Hinton(2010)}]{nairRectifiedLinearUnits}
\bibinfo{author}{Nair, V.}, \bibinfo{author}{Hinton, G.E.},
  \bibinfo{year}{2010}.
\newblock \bibinfo{title}{Rectified {{Linear Units Improve Restricted Boltzmann
  Machines}}} , \bibinfo{pages}{8}.
%Type = Article
\bibitem[{Narciso and Martins(2020)}]{narcisoApplicationMachineLearning2020}
\bibinfo{author}{Narciso, D.A.C.}, \bibinfo{author}{Martins, F.G.},
  \bibinfo{year}{2020}.
\newblock \bibinfo{title}{Application of machine learning tools for energy
  efficiency in industry: {{A}} review}.
\newblock \bibinfo{journal}{Energy Reports} \bibinfo{volume}{6},
  \bibinfo{pages}{1181--1199}.
\newblock \DOIprefix\doi{10/gnmcpk}.
%Type = Incollection
\bibitem[{Paszke et~al.(2019)Paszke, Gross, Massa, Lerer, Bradbury, Chanan,
  Killeen, Lin, Gimelshein, Antiga, Desmaison, Kopf, Yang, DeVito, Raison,
  Tejani, Chilamkurthy, Steiner, Fang, Bai and Chintala}]{NEURIPS2019_9015}
\bibinfo{author}{Paszke, A.}, \bibinfo{author}{Gross, S.},
  \bibinfo{author}{Massa, F.}, \bibinfo{author}{Lerer, A.},
  \bibinfo{author}{Bradbury, J.}, \bibinfo{author}{Chanan, G.},
  \bibinfo{author}{Killeen, T.}, \bibinfo{author}{Lin, Z.},
  \bibinfo{author}{Gimelshein, N.}, \bibinfo{author}{Antiga, L.},
  \bibinfo{author}{Desmaison, A.}, \bibinfo{author}{Kopf, A.},
  \bibinfo{author}{Yang, E.}, \bibinfo{author}{DeVito, Z.},
  \bibinfo{author}{Raison, M.}, \bibinfo{author}{Tejani, A.},
  \bibinfo{author}{Chilamkurthy, S.}, \bibinfo{author}{Steiner, B.},
  \bibinfo{author}{Fang, L.}, \bibinfo{author}{Bai, J.},
  \bibinfo{author}{Chintala, S.}, \bibinfo{year}{2019}.
\newblock \bibinfo{title}{Pytorch: An imperative style, high-performance deep
  learning library}, in: \bibinfo{editor}{Wallach, H.},
  \bibinfo{editor}{Larochelle, H.}, \bibinfo{editor}{Beygelzimer, A.},
  \bibinfo{editor}{d\textquotesingle Alch\'{e}-Buc, F.}, \bibinfo{editor}{Fox,
  E.}, \bibinfo{editor}{Garnett, R.} (Eds.), \bibinfo{booktitle}{Advances in
  Neural Information Processing Systems 32}. \bibinfo{publisher}{Curran
  Associates, Inc.}, pp. \bibinfo{pages}{8024--8035}.
\newblock \URLprefix
  \url{http://papers.neurips.cc/paper/9015-pytorch-an-imperative-style-high-performance-deep-learning-library.pdf}.
%Type = Article
\bibitem[{Pedregosa et~al.(2011)Pedregosa, Varoquaux, Gramfort, Michel,
  Thirion, Grisel, Blondel, Prettenhofer, Weiss, Dubourg
  et~al.}]{pedregosa2011scikit}
\bibinfo{author}{Pedregosa, F.}, \bibinfo{author}{Varoquaux, G.},
  \bibinfo{author}{Gramfort, A.}, \bibinfo{author}{Michel, V.},
  \bibinfo{author}{Thirion, B.}, \bibinfo{author}{Grisel, O.},
  \bibinfo{author}{Blondel, M.}, \bibinfo{author}{Prettenhofer, P.},
  \bibinfo{author}{Weiss, R.}, \bibinfo{author}{Dubourg, V.}, et~al.,
  \bibinfo{year}{2011}.
\newblock \bibinfo{title}{Scikit-learn: Machine learning in python}.
\newblock \bibinfo{journal}{the Journal of machine Learning research}
  \bibinfo{volume}{12}, \bibinfo{pages}{2825--2830}.
%Type = Misc
\bibitem[{Ramachandran et~al.(2017)Ramachandran, Zoph and
  Le}]{ramachandranSearchingActivationFunctions2017}
\bibinfo{author}{Ramachandran, P.}, \bibinfo{author}{Zoph, B.},
  \bibinfo{author}{Le, Q.V.}, \bibinfo{year}{2017}.
\newblock \bibinfo{title}{Searching for {{Activation Functions}}}.
\newblock \DOIprefix\doi{10.48550/arXiv.1710.05941},
  \href{http://arxiv.org/abs/1710.05941}{{\tt arXiv:1710.05941}}.
%Type = Article
\bibitem[{Rana et~al.(2019)Rana, Dey, Tiberi, Vispa, Raspa, Duranti, Ghavami
  and Dudley}]{ranaMachineLearningApproaches}
\bibinfo{author}{Rana, s.}, \bibinfo{author}{Dey, M.}, \bibinfo{author}{Tiberi,
  G.}, \bibinfo{author}{Vispa, A.}, \bibinfo{author}{Raspa, G.},
  \bibinfo{author}{Duranti, M.}, \bibinfo{author}{Ghavami, M.},
  \bibinfo{author}{Dudley, S.}, \bibinfo{year}{2019}.
\newblock \bibinfo{title}{Machine {{Learning Approaches}} for {{Automated
  Lesion Detection}} in {{Microwave Breast Imaging Clinical Data}}} ,
  \bibinfo{pages}{12}.
%Type = Article
\bibitem[{Roshani et~al.(2021)Roshani, Jamshidi, Mohebi and
  Roshani}]{roshaniDesignModelingCompact2021}
\bibinfo{author}{Roshani, S.}, \bibinfo{author}{Jamshidi, M.B.},
  \bibinfo{author}{Mohebi, F.}, \bibinfo{author}{Roshani, S.},
  \bibinfo{year}{2021}.
\newblock \bibinfo{title}{Design and {{Modeling}} of a {{Compact Power
  Divider}} with {{Squared Resonators Using Artificial Intelligence}}}.
\newblock \bibinfo{journal}{Wireless Personal Communications}
  \bibinfo{volume}{117}, \bibinfo{pages}{2085--2096}.
\newblock \DOIprefix\doi{10.1007/s11277-020-07960-5}.
%Type = Article
\bibitem[{Rumelhart et~al.(1986)Rumelhart, Hinton and
  Williams}]{rumelhartLearningRepresentationsBackpropagating1986}
\bibinfo{author}{Rumelhart, D.E.}, \bibinfo{author}{Hinton, G.E.},
  \bibinfo{author}{Williams, R.J.}, \bibinfo{year}{1986}.
\newblock \bibinfo{title}{Learning representations by back-propagating errors}.
\newblock \bibinfo{journal}{Nature} \bibinfo{volume}{323},
  \bibinfo{pages}{533--536}.
\newblock \DOIprefix\doi{10/cvjdpk}.
%Type = Article
\bibitem[{Sarroff(2018)}]{sarroffComplexNeuralNetworks}
\bibinfo{author}{Sarroff, A.M.}, \bibinfo{year}{2018}.
\newblock \bibinfo{title}{Complex {{Neural Networks}} for {{Audio}}} ,
  \bibinfo{pages}{115}.
%Type = Article
\bibitem[{Schonlau and Zou(2020)}]{schonlauRandomForestAlgorithm2020}
\bibinfo{author}{Schonlau, M.}, \bibinfo{author}{Zou, R.Y.},
  \bibinfo{year}{2020}.
\newblock \bibinfo{title}{The random forest algorithm for statistical
  learning}.
\newblock \bibinfo{journal}{The Stata Journal} \bibinfo{volume}{20},
  \bibinfo{pages}{3--29}.
\newblock \DOIprefix\doi{10/ggqh2z}.
%Type = Misc
\bibitem[{Simonyan and Zisserman(2015)}]{simonyanVeryDeepConvolutional2015}
\bibinfo{author}{Simonyan, K.}, \bibinfo{author}{Zisserman, A.},
  \bibinfo{year}{2015}.
\newblock \bibinfo{title}{Very {{Deep Convolutional Networks}} for
  {{Large-Scale Image Recognition}}}.
\newblock \href{http://arxiv.org/abs/1409.1556}{{\tt arXiv:1409.1556}}.
%Type = Article
\bibitem[{Srivastava et~al.(2014)Srivastava, Hinton, Krizhevsky, Sutskever and
  Salakhutdinov}]{srivastavaDropoutSimpleWay2014}
\bibinfo{author}{Srivastava, N.}, \bibinfo{author}{Hinton, G.},
  \bibinfo{author}{Krizhevsky, A.}, \bibinfo{author}{Sutskever, I.},
  \bibinfo{author}{Salakhutdinov, R.}, \bibinfo{year}{2014}.
\newblock \bibinfo{title}{Dropout: {{A Simple Way}} to {{Prevent Neural
  Networks}} from {{Overfitting}}}.
\newblock \bibinfo{journal}{Journal of Machine Learning Research}
  \bibinfo{volume}{15}, \bibinfo{pages}{1929--1958}.
%Type = Misc
\bibitem[{Srivastava et~al.(2015)Srivastava, Greff and
  Schmidhuber}]{srivastavaHighwayNetworks2015}
\bibinfo{author}{Srivastava, R.K.}, \bibinfo{author}{Greff, K.},
  \bibinfo{author}{Schmidhuber, J.}, \bibinfo{year}{2015}.
\newblock \bibinfo{title}{Highway {{Networks}}}.
\newblock \href{http://arxiv.org/abs/1505.00387}{{\tt arXiv:1505.00387}}.
%Type = Article
\bibitem[{Supriya(2020)}]{supriyaTriggerWordRecognition2020}
\bibinfo{author}{Supriya, K.}, \bibinfo{year}{2020}.
\newblock \bibinfo{title}{Trigger {{Word Recognition}} using {{LSTM}}}.
\newblock \bibinfo{journal}{International Journal of Engineering Research and}
  \bibinfo{volume}{V9}.
\newblock \DOIprefix\doi{10/gp4prm}.
%Type = Article
\bibitem[{Trabelsi et~al.(2018)Trabelsi, Bilaniuk, Zhang, Serdyuk, Subramanian,
  Jo{\textbackslash}{\textasciitilde}ao, Mehri, Rostamzadeh, Bengio and
  Christopher}]{trabelsiDeepComplexNetworks2018}
\bibinfo{author}{Trabelsi, C.}, \bibinfo{author}{Bilaniuk, O.},
  \bibinfo{author}{Zhang, Y.}, \bibinfo{author}{Serdyuk, D.},
  \bibinfo{author}{Subramanian, S.},
  \bibinfo{author}{Jo{\textbackslash}{\textasciitilde}ao},
  \bibinfo{author}{Mehri, S.}, \bibinfo{author}{Rostamzadeh, N.},
  \bibinfo{author}{Bengio, Y.}, \bibinfo{author}{Christopher},
  \bibinfo{year}{2018}.
\newblock \bibinfo{title}{Deep {{Complex Networks}}}.
\newblock \bibinfo{journal}{arXiv pre-print server}
  \DOIprefix\doi{10.48550/arxiv.1705.09792}.
%Type = Article
\bibitem[{Travassos et~al.(2020)Travassos, Avila and
  Ida}]{travassosArtificialNeuralNetworks2020}
\bibinfo{author}{Travassos, X.L.}, \bibinfo{author}{Avila, S.L.},
  \bibinfo{author}{Ida, N.}, \bibinfo{year}{2020}.
\newblock \bibinfo{title}{Artificial {{Neural Networks}} and {{Machine
  Learning}} techniques applied to {{Ground Penetrating Radar}}: {{A}} review}.
\newblock \bibinfo{journal}{Applied Computing and Informatics}
  \bibinfo{volume}{17}, \bibinfo{pages}{296--308}.
\newblock \DOIprefix\doi{10.1016/j.aci.2018.10.001}.
%Type = Article
\bibitem[{Vaswani et~al.(2017)Vaswani, Shazeer, Parmar, Uszkoreit, Jones,
  Gomez, Kaiser and Polosukhin}]{vaswaniAttentionAllYou2017}
\bibinfo{author}{Vaswani, A.}, \bibinfo{author}{Shazeer, N.},
  \bibinfo{author}{Parmar, N.}, \bibinfo{author}{Uszkoreit, J.},
  \bibinfo{author}{Jones, L.}, \bibinfo{author}{Gomez, A.N.},
  \bibinfo{author}{Kaiser, {\L}.}, \bibinfo{author}{Polosukhin, I.},
  \bibinfo{year}{2017}.
\newblock \bibinfo{title}{Attention is all you need}.
\newblock \bibinfo{journal}{Advances in neural information processing systems}
  \bibinfo{volume}{30}.
\newblock \DOIprefix\doi{10.48550/arxiv.1706.03762}.
%Type = Article
\bibitem[{Werbos and John(1974)}]{werbosRegressionNewTools1974}
\bibinfo{author}{Werbos, P.}, \bibinfo{author}{John, P.}, \bibinfo{year}{1974}.
\newblock \bibinfo{title}{Beyond regression : New tools for prediction and
  analysis in the behavioral sciences /} .
%Type = Book
\bibitem[{White(2004)}]{white2004high}
\bibinfo{author}{White, J.F.}, \bibinfo{year}{2004}.
\newblock \bibinfo{title}{High frequency techniques: An introduction to rf and
  microwave design and computer simulation}.
\newblock \bibinfo{publisher}{John Wiley \& Sons}.
%Type = Article
\bibitem[{Yang and Bose(2005)}]{yangLandmineDetectionClassification2005}
\bibinfo{author}{Yang, C.C.}, \bibinfo{author}{Bose, N.}, \bibinfo{year}{2005}.
\newblock \bibinfo{title}{Landmine {{Detection}} and {{Classification With
  Complex-Valued Hybrid Neural Network Using Scattering Parameters Dataset}}}.
\newblock \bibinfo{journal}{IEEE Transactions on Neural Networks}
  \bibinfo{volume}{16}, \bibinfo{pages}{743--753}.
\newblock \DOIprefix\doi{10.1109/TNN.2005.844906}.
%Type = Inproceedings
\bibitem[{Zeiler et~al.(2013)Zeiler, Ranzato, Monga, Mao, Yang, Le, Nguyen,
  Senior, Vanhoucke, Dean and Hinton}]{zeilerRectifiedLinearUnits2013}
\bibinfo{author}{Zeiler, M.}, \bibinfo{author}{Ranzato, M.},
  \bibinfo{author}{Monga, R.}, \bibinfo{author}{Mao, M.},
  \bibinfo{author}{Yang, K.}, \bibinfo{author}{Le, Q.},
  \bibinfo{author}{Nguyen, P.}, \bibinfo{author}{Senior, A.},
  \bibinfo{author}{Vanhoucke, V.}, \bibinfo{author}{Dean, J.},
  \bibinfo{author}{Hinton, G.}, \bibinfo{year}{2013}.
\newblock \bibinfo{title}{On rectified linear units for speech processing}, in:
  \bibinfo{booktitle}{2013 {{IEEE International Conference}} on {{Acoustics}},
  {{Speech}} and {{Signal Processing}}}, \bibinfo{publisher}{{IEEE}},
  \bibinfo{address}{{Vancouver, BC, Canada}}. pp. \bibinfo{pages}{3517--3521}.
\newblock \DOIprefix\doi{10.1109/ICASSP.2013.6638312}.
%Type = Article
\bibitem[{Zhang et~al.(2017)Zhang, Wang, Xu and
  Jin}]{zhangComplexValuedConvolutionalNeural2017}
\bibinfo{author}{Zhang, Z.}, \bibinfo{author}{Wang, H.}, \bibinfo{author}{Xu,
  F.}, \bibinfo{author}{Jin, Y.Q.}, \bibinfo{year}{2017}.
\newblock \bibinfo{title}{Complex-{{Valued Convolutional Neural Network}} and
  {{Its Application}} in {{Polarimetric SAR Image Classification}}}.
\newblock \bibinfo{journal}{IEEE Transactions on Geoscience and Remote Sensing}
  \bibinfo{volume}{55}, \bibinfo{pages}{7177--7188}.
\newblock \DOIprefix\doi{10/gcqg44}.

\end{thebibliography}

%% else use the following coding to input the bibitems directly in the
%% TeX file.

% \begin{thebibliography}{00}

% %% \bibitem[Author(year)]{label}
% %% Text of bibliographic item

% \bibitem[ ()]{}

% \end{thebibliography}
\end{document}